\documentclass[reprint,pre,twocolumns]{revtex4-2}
\usepackage{graphicx}
\usepackage{amssymb}
\usepackage{physics}
\usepackage{xcolor}
\definecolor{ForestGreen}{RGB}{34,139,34}
\usepackage{soul,ulem}
\graphicspath{ {./images/} }

\def\avk#1{\textcolor{black}{#1}}

\begin{document}

\title{Collisions of localized patterns in a nonvariational Swift-Hohenberg equation}

\author{Mathi Raja} \author{Adrian van Kan}
\author{Benjamin Foster}
\author{Edgar Knobloch}
\affiliation{Department of Physics, University of California at Berkeley, Berkeley, California 94720, USA}

\date{\today}

\begin{abstract}

The cubic-quintic Swift-Hohenberg equation (SH35) has been proposed as an order parameter description of several convective systems with reflection symmetry in the layer midplane, including binary fluid convection. We use numerical continuation, together with extensive direct numerical simulations (DNSs), to study SH35 with an additional nonvariational quadratic term to model the effects of breaking the midplane reflection symmetry. The nonvariational structure of the model leads to the propagation of asymmetric spatially localized structures (LSs). An asymptotic prediction for the drift velocity of such structures, derived in the limit of weak symmetry breaking, is validated numerically. Next, we present an extensive study of possible collision scenarios between identical and nonidentical traveling structures, varying a temperature-like control parameter. These collisions are inelastic and result in stationary or traveling structures. Depending on system parameters and the types of structures colliding, the final state may be a simple bound state of the initial LSs, but it can also be longer or shorter than the sum of the two initial states as a result of nonlinear interactions. The Maxwell point of the variational system, where the free energy of the global pattern state equals that of the trivial state, is shown to have no bearing on which of these scenarios is realized. Instead, we argue that the stability properties of bound states are key. While individual LSs lie on a modified snakes-and-ladders structure in the nonvariational SH35, the multi-pulse bound states resulting from collisions lie on isolas in parameter space, disconnected from the trivial solution. In the gradient SH35, such isolas are always of figure-eight shape, but in the present non-gradient case they are generically more complex, although the figure-eight shape is preserved in a small subset of cases. Some of these complex isolas are shown to terminate in T-point bifurcations. A reduced model is proposed to describe the interactions between the tails of the LSs. The model consists of two coupled ordinary differential equations (ODEs) capturing the oscillatory structure of SH35 at the linear level. It contains three parameters: two interaction amplitudes and a phase, whose values are deduced from high-resolution DNSs using gradient descent optimization. For collisions leading to the formation of simple bound states, the reduced model reproduces the trajectories of LSs with high quantitative accuracy. When nonlinear interactions lead to the creation or deletion of wavelengths the model performs less well. Finally, we propose an effective signature of a given interaction in terms of net attraction or repulsion relative to free propagation. It is found that interactions can be attractive or repulsive in the net, irrespective of whether the two closest interacting extrema are of the same or opposite signs. Our findings highlight the rich temporal dynamics described by this bistable nonvariational SH35, and show that the interactions in this system can be quantitatively captured, to a significant extent, by a highly reduced ODE model.
\end{abstract}

\maketitle

%%%%%%%%%%%%%%%%%%%%%%%%%%%%%%%%%%%%%%%%%%%%%%%%%%%%%%%%%%%
%%%%%%%%%%%%%%%%%%%%%%%%%%%%%%%%%%%%%%%
\section{Introduction}
\label{sec:intro}
%%%%%%%%%%%%%%%%%%%%%%%%%%%%%%%%%%%%%%
{Spatially localized structures (LSs) are observed in a wide variety of physical systems, from solitary water waves \cite{korteweg1895xli} to neurons \cite{laing2002multiple}, fluid convection \cite{ghorayeb1997double,batiste2006spatially}, shear flows \cite{gad1981growth,schneider2010localized} and reaction-diffusion systems \cite{coullet2000stable,yochelis2008formation}, to name only a few. Generically, these systems are subject to dissipation and require forcing to maintain the structure, see \cite{knobloch2015spatial} for a review of spatial localization in such systems. A simple model of pattern formation in forced dissipative systems is provided by the bistable Swift-Hohenberg equation, originally suggested in the context of pattern formation in Rayleigh-Bénard convection \cite{hohenberg1992effects,ma2011diagrammatic}. This equation supports well-known localized solutions that are organized in a \textit{snakes-and-ladders} bifurcation structure \cite{burke2006localized,burke2007snakes,burke2007homoclinic}.}

{When the Swift-Hohenberg equation has gradient structure, solutions with nontrivial time dependence are precluded. However, nongradient generalizations of the Swift-Hohenberg equation arise frequently in applications \cite{kozyreff2007nonvariational} and these permit both time dependence, see e.g. \cite{burke2012localized}, and persistent propagation, e.g. \cite{burke2009swift}. In this paper, we consider a specific instance of such models, namely the one-dimensional cubic-quintic Swift-Hohenberg equation with broken reflection symmetry,} 
\begin{equation}
    \partial_t u = ru - (1+\partial_x^2)^2 u + b_3 u^3- u^5 + \epsilon (\partial_x u)^2. \label{eq:she_general}
\end{equation}
{Here the parameter $\epsilon$ controls both the nongradient structure of the equation (the equation has gradient structure when $\epsilon=0$) and the breaking of the symmetry $u\rightarrow -u$ (the equation is invariant under $u\rightarrow -u$ when $\epsilon=0$). Since the equation is also symmetric under spatial reflections, $x\rightarrow -x$, both effects are required for spontaneous propagation of LSs in this system.}

{Equation~(\ref{eq:she_general}) was suggested in \cite{houghton2011swift} as a model of binary fluid convection with broken midplane reflection symmetry and its properties are indeed in qualitative agreement with direct numerical simulations of the Navier-Stokes equations describing this system \cite{mercader2013travelling}. The present work extends significantly the collision studies undertaken in \cite{houghton2011swift} and clarifies a number of key issues.}
%%%%%%%%%%%%%%%%%%%%%%%%%%%%%%%%%%%%%%%%%%%%%%%%%%%%%%%%%%
\begin{figure*}
\includegraphics[width=\textwidth]{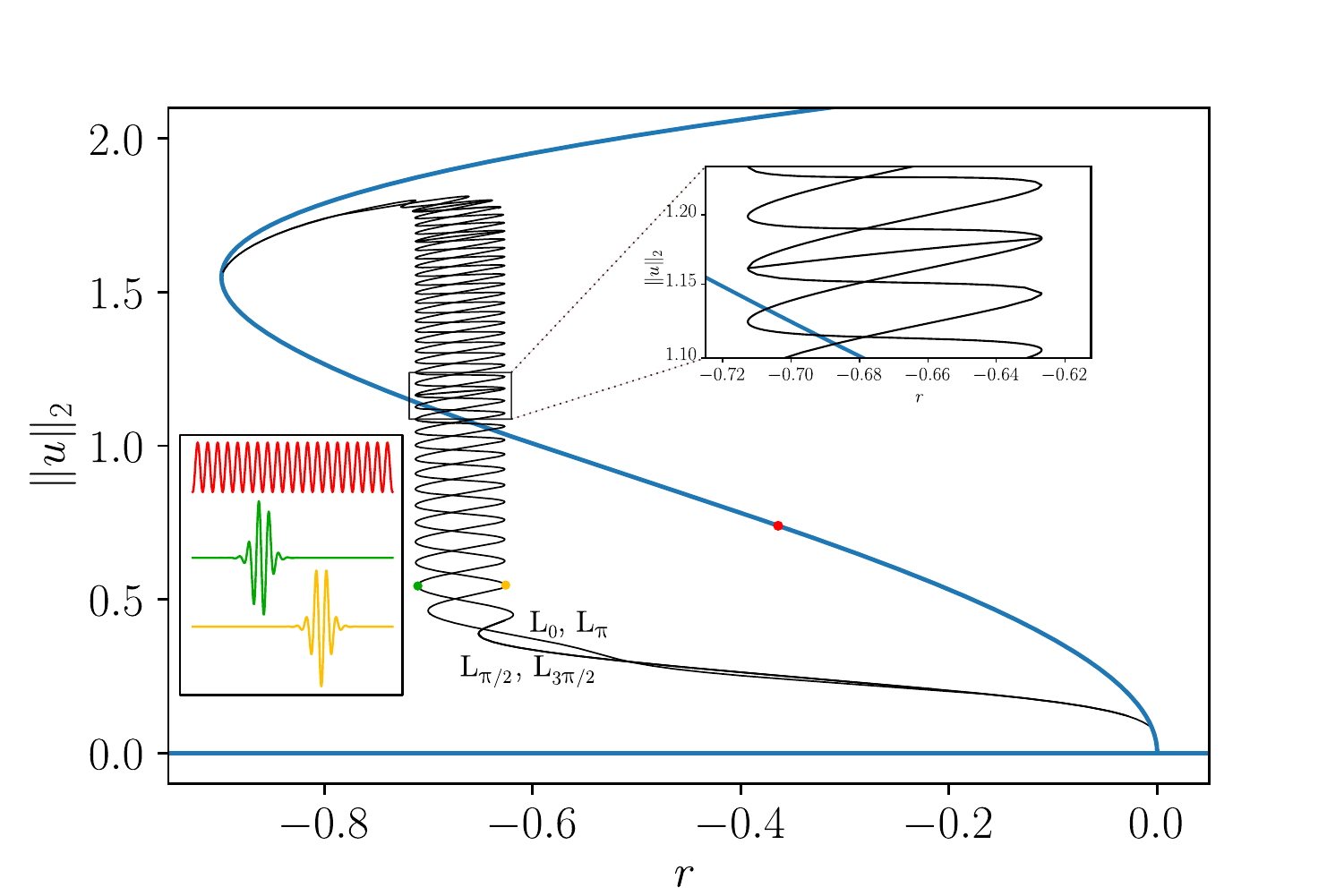}
\caption{Bifurcation diagram for Eq.~(\ref{eq:she_general}) with $\epsilon = 0$.  The patterned state branch is shown in blue with a sample solution profile in red (lower left inset).  The snaking branches of symmetric ($L_0, L_{\pi}$) and antisymmetric ($L_{\pi/2}, L_{3\pi/2}$) LSs are shown in black (sample profiles in yellow and green, respectively). For clarity, only three of the interconnecting rung states are shown (detail in upper right inset), cf. \cite{houghton2011swift}. Larger norm indicates longer LSs. Stable solutions are found on branches with positive slope.}
\label{fig:eps_0_bifurcation}
\end{figure*}
%%%%%%%%%%%%%%%%%%%%%%%%%%%%%%%%%%%%%%%%%%%%%%%%%%%%%%%%%%
{In the following, we refer to Eq.~(\ref{eq:she_general}) as SH35. When $\epsilon=0$, the problem reduces to variational form such that, on a domain of any size $L$, there exists a free energy functional
\begin{equation}
    \mathcal{F}[u(x)] = \int_0^L \left(- \frac{1}{2}r u^2 +  \frac{1}{2} [(1+\partial_x^2)u]^2 - \frac{b_3}{4}u^4 + \frac{u^6}{6}\right) dx, \label{eq:free_energy}
\end{equation}
with the property that $\partial_t u = - \frac{\delta\mathcal{F}}{\delta u}$. Thus $\mathcal{F}$ decreases along trajectories towards local minima as $t\rightarrow\infty$, and these represent stable steady states of the system. The free energy of spatially periodic patterns passes through zero at a $r=r_M$, known as the \textit{Maxwell point}. In the vicinity of the  Maxwell point one can create a variety of localized structures involving both the pattern state and the trivial state $u=0$ at little or no cost in energy. These are of two types, localized even solutions, denoted here as $L_0$ ($L_\pi$) if their maximum (minimum) is located at their center, and localized odd solutions, denoted as $L_{\frac{\pi}{2}}$ ($L_{\frac{3\pi}{2}}$) if they have a negative (positive) slope at the center.

When $0<\epsilon\ll1$, the symmetry $u\to - u$ as well as the variational structure is broken but similar solutions continue to exist, albeit with modified properties, as described in \cite{houghton2011swift}. The symmetric solution branches $L_0,L_\pi$ remain symmetric and stationary. Odd solution profiles cease to be odd and hence propagate. In this paper, we first study the propagation of isolated structures and then go on to investigate in detail the collisions that can result. In contrast to the collisions familiar from studies of integrable partial differential equations on the real line, here the collisions are inelastic and can lead to annihilation and sticking as well as scattering. One question of interest concerns the role, if any, played by the Maxwell point of the variational system in such collisions when $\epsilon$ is small: for example, is the collision process accompanied by nucleation or annihilation of new wavelengths according to the free energy minimization principle valid at $\epsilon=0$ or, if this principle is not followed, what other mechanism determines collision outcomes?

%Initial results on such collisions were already presented in \cite{houghton2011swift}, and it was suggested that they might be relevant to collisions of convectons in binary fluid convection \cite{mercader2013travelling}, but no detailed investigation of traveling LSs in the nonvariational SH35 has been done to date. Here, we address this gap in understanding by systematically studying the propagation and collisions of different LSs and the dependence of this process on control parameters.

The remainder of this paper is structured as follows. In section \ref{sec:bif_ana}, we describe the general bifurcation structure of Eq.~(\ref{eq:she_general}) obtained from numerical continuation. Next, in section \ref{sec:drift_speed} we present an asymptotic computation of the drift speed of asymmetric LS in the limit of weak symmetry breaking $0<\epsilon \ll1$, whose accuracy is confirmed by comparison with direct numerical simulations (DNSs) and numerical continuation. In section \ref{sec:overview_collisions}, we present the results of extensive DNSs of all possible collision scenarios in this system, and describe the dependence of the collision outcome on the control parameter $r$, with the stability of multi-pulse bound states playing a key role. We also show that the bound-state solutions arising from collisions are in many cases a part of non-trivial isolas, whose structure depends on the symmetry breaking parameter $\epsilon$. In section \ref{sec:reduced_model}, we present a reduced model of the interactions between colliding patterns, which is based on the linear structure of SH35 and accurately captures the trajectories of LSs so long as no significant nonlinear interactions creating or destroying wavelengths occur. The paper concludes in section \ref{sec:conclusions} with a discussion of our results. All our results are obtained for the choice $b_3=2$ as used in \cite{houghton2011swift}, employing periodic boundary conditions on a domain of size $L=40\pi$.

%%%%%%%%%%%%%%%%%%%%%%%%%%%%%%%%%55
\begin{figure}
    \centering
    \includegraphics[width=0.5\textwidth]{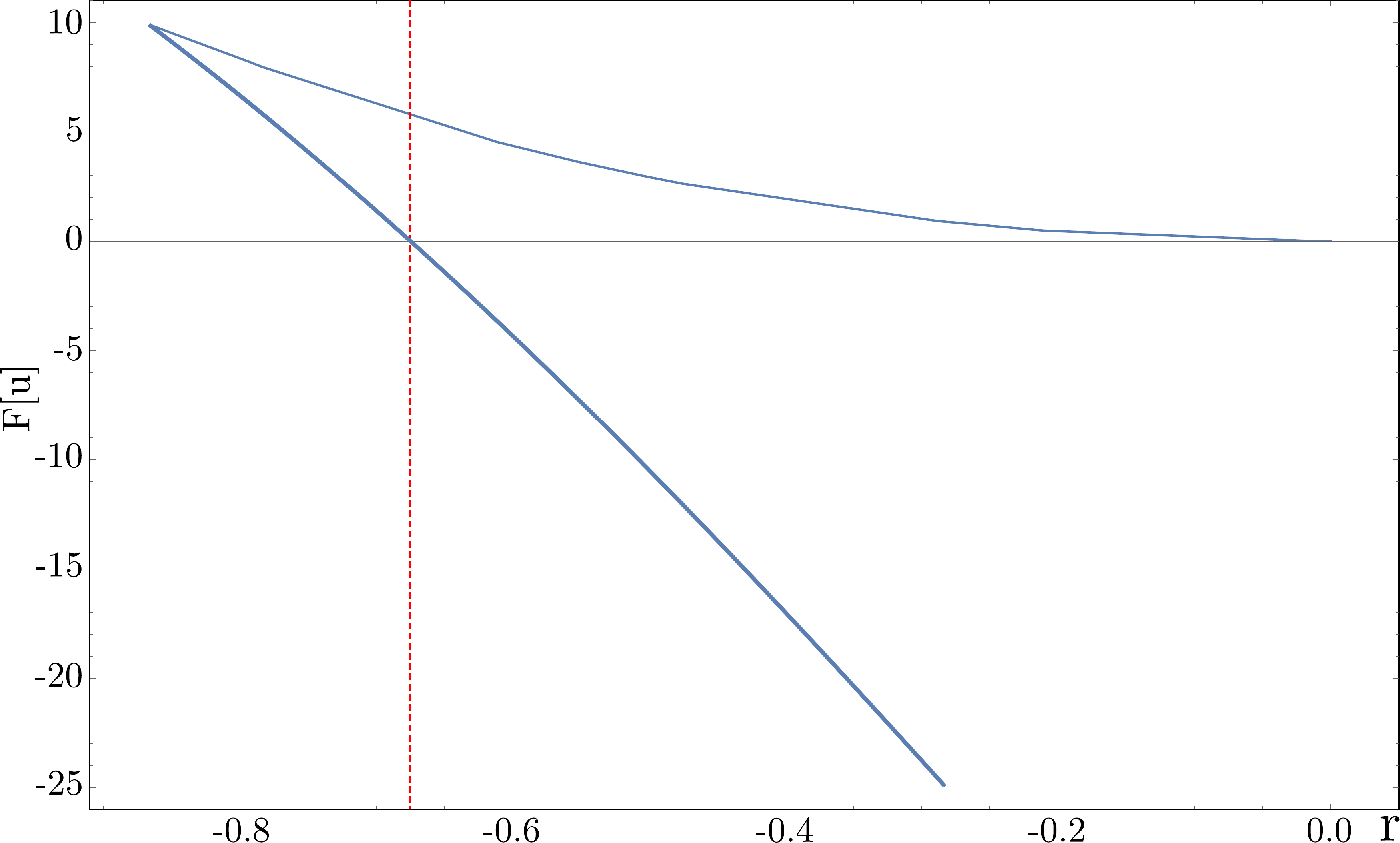}
    \caption{Free energy $F[u]$, defined in Eq.~(\ref{eq:free_energy}), of the periodic pattern state versus $r$. Thick (thin) lines correspond to stable (unstable) solutions. The dotted vertical red line indicates the Maxwell point $r_M\approx -0.675$, where the free energy changes sign.}
    \label{fig:free_energy}
\end{figure}
%%%%%%%%%%%%%%%%%%%%%%%%%%%%%%%%%%%5

%%%%%%%%%%%%%%%%%%%%%%%%%%%%%%%%%%%%%%%%%%%%%%%%%%%%%%
\begin{figure*}
\includegraphics[width=\textwidth]{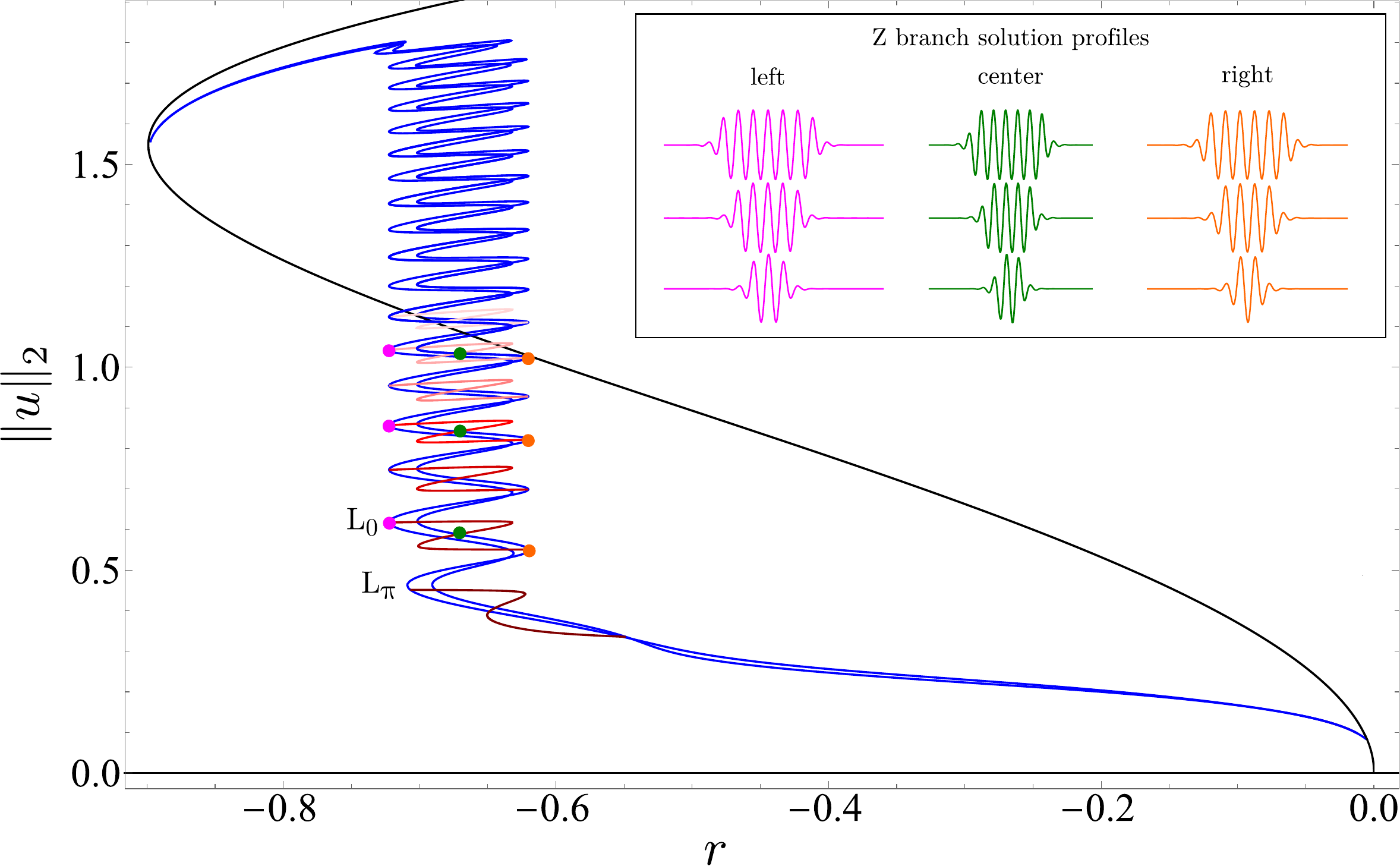}
\caption{Bifurcation diagram for Eq.~(\ref{eq:she_general}) with $\epsilon = 0.03$.  Stable Z branch states with at least two wavelengths exist in the range $-0.70 \lesssim r \lesssim -0.63$.  Inset: sample solution profiles at color-coded locations in ascending order, cf. \cite{houghton2011swift}. Larger norm indicates longer LSs. Stable solutions are found on branches with positive slope.}
\label{fig:eps_03_bif}
\end{figure*}
%%%%%%%%%%%%%%%%%%%%%%%%%%%%%%%%%%%%%%%%%%5

%%%%%%%%%%%%%%%%%%%%%%%%%%%%%%%%%%%%%%%%%%%%
\section{Bifurcation analysis} 
\label{sec:bif_ana}
%%%%%%%%%%%%%%%%%%%%%%%%%%%%%%%%%%%%%%%%%%%%
In this section we provide an overview of the solution structure of Eq.~(\ref{eq:she_general}), first in the case $\epsilon=0$, and then in the case $\epsilon>0$, obtained from numerical continuation using AUTO \cite{doedel08auto-07p} and pde2path \cite{uecker_numerical_2021}. The results are presented in terms of the $L_2$ norm of the solutions $u(x)$ given by
\begin{equation}
    \|u\|_2 \equiv \sqrt{\frac{1}{L} \int_0^L u^2(x) dx}
    \label{eq:def_L2}
\end{equation}
and provide important background information for subsequent sections.

%%%%%%%%%%%%%%%%%%%%%%%55
\subsection{The variational case: $\epsilon=0$}
%%%%%%%%%%%%%%%%%%%%%%%%%%%%%%%%%5
We first consider the variational problem with $\epsilon=0$ and employ numerical continuation to construct the bifurcation diagram shown in Fig.~\ref{fig:eps_0_bifurcation}, cf. \cite{burke2007snakes}. A trivial flat state $u=0$ exists at all $r$, and undergoes a subcritical bifurcation to a periodic pattern at $r=0$ (red profile in lower left inset in Fig.~\ref{fig:eps_0_bifurcation}). Four snaking branches bifurcate from the periodic pattern branch in secondary bifurcations, corresponding to the symmetric LSs $L_0$, $L_{\pi}$ and the antisymmetric LSs $L_{\pi/2}$, $L_{3\pi/2}$ mentioned in the introduction. The branches overlap in pairs owing to the symmetry $u\to-u$, and both sets display characteristic snaking behavior in the vicinity of the Maxwell point $r_M$ \cite{burke2007snakes,knobloch2015spatial}. In addition, rung states connect the symmetric and antisymmetric snaking branches, arising in pitchfork bifurcations close to every saddle-node bifurcation on these branches \cite{houghton2011swift}. Each rung actually corresponds to four branches of unstable asymmetric localized solutions, related by the symmetries $x\to -x$ and $u\to -u$ \cite{burke2007homoclinic}.
%($L_{\pi/2}$ and $L_{3\pi/2}$) with odd symmetry. \blf{I believe this is not correct}
Examples of symmetric and antisymmetric solution profiles are shown in the lower left inset in Fig.~\ref{fig:eps_0_bifurcation}, together with the periodic pattern state. When the LSs fill the available domain the snaking branches reconnect to the periodic pattern.

Figure \ref{fig:free_energy} shows the free energy of the periodic pattern from Fig.~\ref{fig:eps_0_bifurcation} versus $r$. The stable part of the periodic pattern state (thick blue line) has a free energy which decreases monotonically with $r$ and changes sign at the Maxwell point $r_M\approx-0.675$.

%%%%%%%%%%%%%%%%%%%%%%%%%%%%%%%%%%%%%%%%5
\subsection{The nonvariational case: $\epsilon >0$}
%%%%%%%%%%%%%%%%%%%%%%%%%%%%%%%%%%%
Here we describe how the bifurcation structure changes in the nonvariational case, specifically for $\epsilon=0.03$ (Fig.~\ref{fig:eps_03_bif}). As in the case $\epsilon = 0$, there is a trivial branch $u=0$, a periodic pattern branch emerging subcritically from it, and two snaking branches. However, the two snaking branches in Fig.~\ref{fig:eps_03_bif} now correspond to $L_0$ and $L_\pi$, since these states are no longer related by symmetry, and consequently snake in phase before reconnecting to the periodic state. Moreover, the $L_{\pi/2}$, $L_{3\pi/2}$ states and the rung states reconnect, forming a sequence of 'Z'-shaped branches consisting of asymmetric solutions.

As a consequence of the nonvariational structure of the problem when $\epsilon>0$, these asymmetric solutions drift and hence may collide. The structure of several Z branch solutions at different $r$ values is shown in the inset in Fig.~\ref{fig:eps_03_bif}. It is important to observe that the Z branch states may be stable or unstable. Stable solutions are present on the ``diagonal'' part of each Z, i.e. in the range $-0.70 \lesssim r \lesssim -0.63$} (except for the lowest branch, which is stable only between $-0.65\lesssim r \lesssim -0.62$). The corresponding drift velocity $c$ depends on $r$, as shown in Fig.~\ref{fig:c_vs_r} for several of the states depicted in Fig.~\ref{fig:eps_03_bif}. In Fig.~\ref{fig:c_vs_r}, the stable part of any given Z branch corresponds to the segment between $-0.70\lesssim r \lesssim -0.63$, where the drift speed is largest. We emphasize that longer asymmetric LSs drift more slowly than short ones.

%%%%%%%%%%%%%%%%%%%%%%%%%%%%%%%%%%%%%%%%%%%%%%%%%%%%%%%%%%%%%%%5
\begin{figure}
    \centering
    \includegraphics[width = 0.5\textwidth]{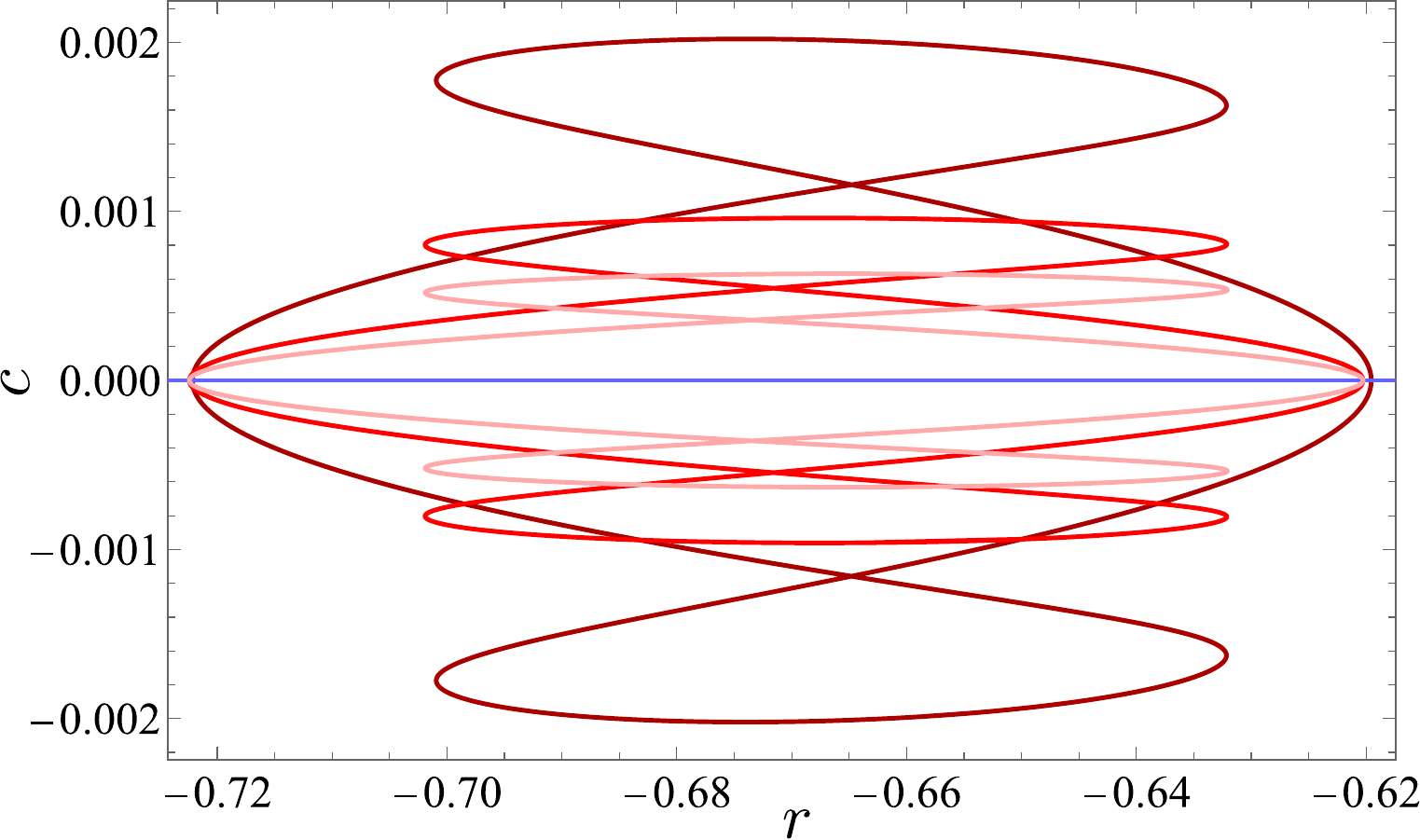}
    \label{fig:r_vs_c}
    \caption{Drift velocity $c$ versus $r$ for various Z branch states at $\epsilon=0.03$; colors correspond to the branches in Fig.~\ref{fig:eps_03_bif}. The curves shown correspond to the Z branches for which profiles are displayed in Fig.~\ref{fig:eps_03_bif}.}
    \label{fig:c_vs_r}
\end{figure}
%%%%%%%%%%%%%%%%%%%%%%%%%%%%%%%%%%%%%%%%%%%%%%%%%%%%%%%%%%%%%%%

%%%%%%%%%%%%%%%%%%%%%%%%%%%%%%%%%%%%%%%%%%%%%%%%%%%%%%%%%%%%%%%
\section{Drift velocity of isolated structures}
\label{sec:drift_speed}
%%%%%%%%%%%%%%%%%%%%%%%%%%%%%%%%%%%%%%%%%%%%%%%%%%%%%%%%%%%%%%%
\subsection{Asymptotic theory}
%%%%%%%%%%%%%%%%%%%%%%%%%%%%%%%%%%%%%%%%%%%%%%%%%%%%%%%%%%%%%%%
To compute the drift velocity of asymmetric LSs using perturbation theory in the limit of weak symmetry breaking, $0<\epsilon \ll 1$, we introduce the fast and slow time variables
\begin{equation}
\tau = \epsilon^{\frac{1}{2}}t, \qquad
T = \epsilon t
\end{equation}
 and denote their spatial phase by $\theta(T)$. Following the calculation presented in \cite{burke2009swift}, we posit the expansion
\begin{align}
u(x,t) = \hphantom{l}&U_0[x-\theta(T)]+\epsilon^{\frac{1}{2}}u_1[x-\theta(T), \tau] \nonumber \\ &+\epsilon u_2[x-\theta(T), \tau] + o(\epsilon),
\label{eq:speed_ansatz}
\end{align}
where $U_0$ is a known moving pattern whose propagation velocity we seek to determine. At leading order, Eq.~(\ref{eq:she_general}) implies 
\begin{equation}
    rU_0-(1+\partial_x^2)^2U_0+b_2U_0^3-U_0^5 = 0, \label{eq:asymp_O_eps_zero}
\end{equation}
while at $O(\epsilon^\frac{1}{2})$ we obtain
\begin{equation}
\mathcal{L} u_1 = ru_1-(1+\partial_x^2)^2u_1+3b_2U_0^2u_1-5U_0^4u_1 = 0, 
\label{eq:asymp_O_eps_one_half}
\end{equation}
where the operator $\mathcal{L} \equiv r-(1+\partial_x^2)^2+3b_2U_0^2-5U_0^4$ is self-adjoint. \newline

Differentiating Eq.~(\ref{eq:asymp_O_eps_zero}) with respect to $x$, one finds that $\mathcal{L}U_0'=0$, i.e. if $U_0$ solves (\ref{eq:asymp_O_eps_zero}), then $U_0'$ solves (\ref{eq:asymp_O_eps_one_half}). In \cite{burke2009swift}, the authors considered $r$ to be asymptotically close to the edge of the snaking interval, with a focus on the dynamics of depinning. This required taking into account two additional solutions which lie in the null space of $\mathcal{L}$: one symmetric and one asymmetric mode. Here, we instead consider asymmetric states which are located within the snakes-and-ladders structure, away from the saddle-nodes of the snaking branches. Consequently $U_0'$ is the only function in the null space of $\mathcal{L}^\dagger =\mathcal{L}$. Since this translation mode is already included in the Ansatz (\ref{eq:speed_ansatz}) we can set $u_1\equiv 0$. Finally, at $O(\epsilon)$, we obtain 
\begin{equation}
   -U'_0\theta_T = \mathcal{L}u_2+ \epsilon (U'_0)^2. \label{eq:asymp_eps_one}
\end{equation}
Multiplying by $U_0'$ and integrating over the domain, using the fact that $\mathcal{L}$ is self-adjoint and that $\mathcal{L}U_0'=0$, we obtain
\begin{equation}
    -\theta_T \int (U'_0)^2 dx = \int U'_0 \mathcal{L} u_2 + \epsilon (U'_0)^3 dx = \int \epsilon (U'_0)^3 dx. 
    \label{eq:solvability_cond}
\end{equation}
Rearranging, we arrive at the drift velocity,
\begin{align}
c\equiv \theta_T  = -\epsilon\frac{\int_0^L \left(U_0'(x)\right)^3 dx}{\int_0^L \left(U_0'(x)\right)^2 dx},
\label{eq:drift_velocity}
\end{align}
a special case of a more general result \cite{makrides2014predicting}.
%for a general nonvariational term (see their Eqs.~(4.1) and (4.2)). While the step of obtaining the solvability condition in Eq.~\ref{eq:solvability_cond} is also taken there, we believe that our presentation complements the succinct exposition given in \cite{makrides2014predicting}, as does our numerical verification  of it (presented below in Sec.~\ref{ssec:num_verif}).

\begin{figure}
\includegraphics[width=0.5\textwidth]{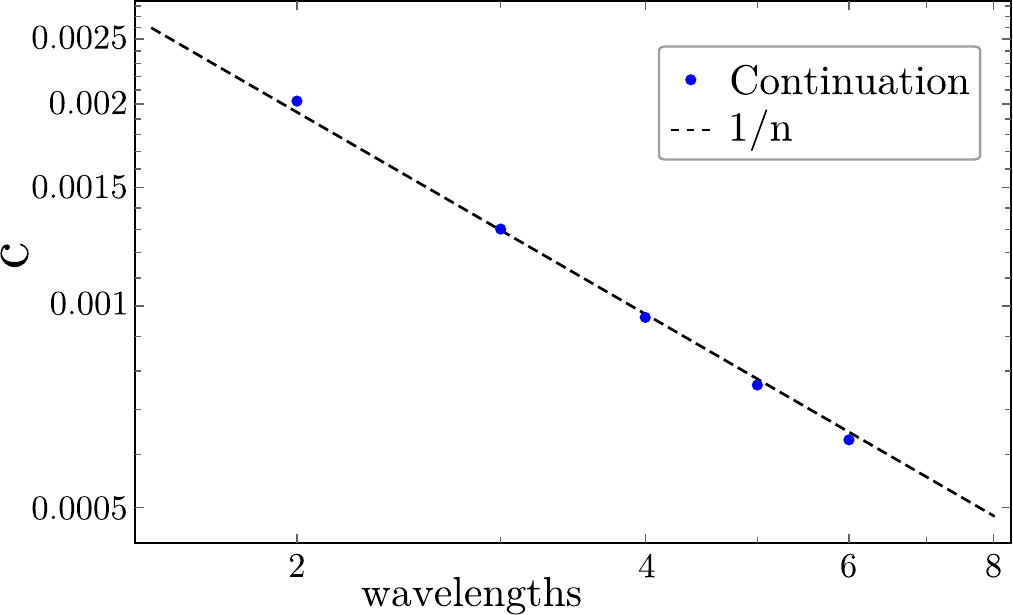}
\caption{Log plot showing $c(r=-0.67)$ (blue) vs. number of wavelengths $n$ overlaid with $1/n$ (black). A satisfactory agreement is observed.}
\label{fig:cpeaks}
\end{figure}

Equation~(\ref{eq:drift_velocity}) is invariant under $(\epsilon,U_0)\to (-\epsilon,-U_0)$, a symmetry that is inherited from Eq.~(\ref{eq:she_general}). For symmetric profiles $U_0(x)$, the numerator vanishes; symmetric states therefore remain at rest even when $\epsilon>0$.
%If $U_0(x)$ is an antisymmetric sinusoidak profile (which is not a solution to Eq. (\ref{eq:she_general})), with $n+1/2$ wavelengths, where $n\in \mathbb{N}$, then the numerator also vanishes in Eq.~(\ref{eq:drift_velocity}).
However, for the asymmetric profiles shown in Fig.~\ref{fig:eps_03_bif} the velocity $c$ is nonzero. This is due to the nonsinusoidal nature of these profiles and in particular the contribution from the fronts at either end of the LS profile. We define the number $n$ of {\it significant extrema} in any given LS as the number of extrema whose amplitude is larger than $1/e$ times the maximum amplitude in the LS, and the number of {\it significant wavelengths} as $1/2$ the number of significant extrema. The adjective \textit{significant} will be dropped in the following when there is no ambiguity. For solutions with many {significant} wavelengths, $n\gg 1$, the numerator converges to a constant, while the denominator grows approximately linearly with $n$. This implies that $c\propto 1/n$ at large $n$, which is quantitatively consistent with the results from numerical continuation, as shown in Fig.~\ref{fig:cpeaks}: longer structures move more slowly, as already highlighted in the discussion of Fig.~\ref{fig:c_vs_r}.
%%%%%%%%%%%%%%%%%%%%%%%%%%%%%%%%%%%%%%%%%%%%%%%%%%%%%%%%%%%%%%%
%%%%%%%%%%%%%%%%%%%%%%%%%%%%%%%%%%%%%%%%%%%%%%%%%%%%%%%%%%%%%%%

%%%%%%%%%%%%%%%%%%%%%%%%%%%%%%%%%%%%%%%%%%%%%%%%%%%%%%%%%%%%%%%
\begin{figure}
\includegraphics[width=0.5\textwidth]{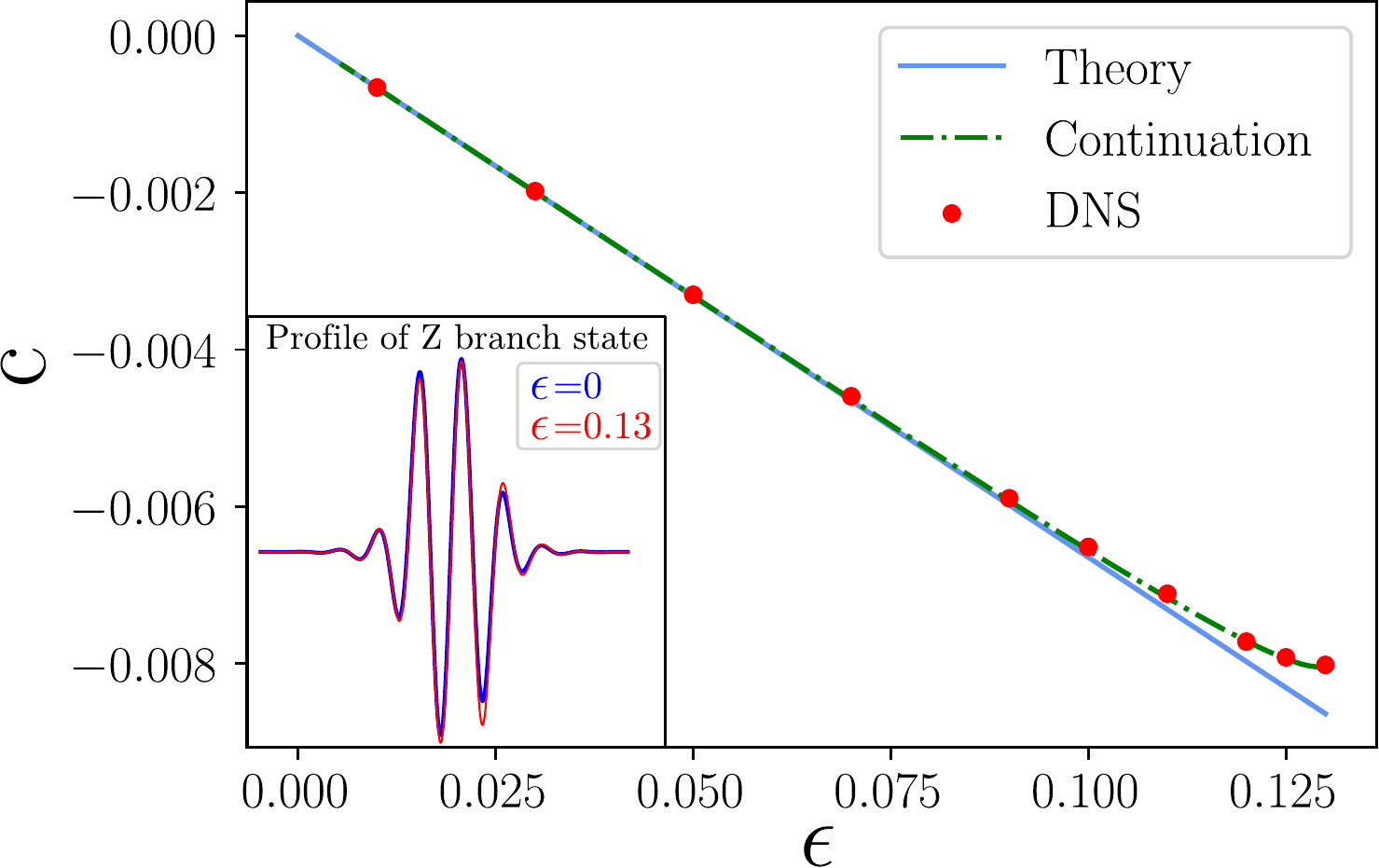}
\caption{Comparison of the theoretically predicted drift velocity $c$ with the corresponding DNS results for stable states on the second lowest Z branch.  Blue solid line indicates the theoretical prediction from Eq.~(\ref{eq:drift_velocity}), with the integrals evaluated numerically. The green dash-dotted line shows the velocity obtained from numerical continuation. Red circles indicate the DNS results. The DNS and numerical continuation are in excellent agreement, including the deviation from asymptotic behavior at larger $\epsilon$. The asymptotic theory is in close quantitative agreement with both DNS and numerical continuation at small $\epsilon$. The Z branch states cease to exist above $\epsilon=0.13$.  Inset shows the Z branch solution profiles at $\epsilon=0$ (blue, antisymmetric) and $\epsilon=0.13$ (red, asymmetric) at $r=-0.66$.}
\label{fig:verify_velocity}
\end{figure}
%%%%%%%%%%%%%%%%%%%%%%%%%%%%%%%%%%%%%%%%%%%%%%%%%%%%%%%%%%%%%%%

%%%%%%%%%%%%%%%%%%%%%%%%%%%%%%%%%%%%%%%%%%%%%%%%%%%%%%%%%%%%%%%
\subsection{Numerical verification}
%%%%%%%%%%%%%%%%%%%%%%%%%%%%%%%%%%%%%%%%%%%%%%%%%%%%%%%%%%%%%%%
\label{ssec:num_verif}
To determine the range of validity of the prediction in Eq.~(\ref{eq:drift_velocity}), we perform direct numerical simulations (DNSs) of Eq.~(\ref{eq:she_general}) for selected values of the parameter $\epsilon$. We also perform numerical continuation in $\epsilon$. The results reported below extend those in \cite{makrides2014predicting} where the asymptotic result was compared with numerical continuation at only one value of $\epsilon$, $|\epsilon|=0.01$.
%The range of validity of the asymptotic predictions in $\epsilon$ was not quantified. Here, we go beyond \cite{makrides2014predicting} and use DNSs and continuation to test the accuracy of the asymptotic results at different values of $\epsilon$.} 
For this purpose, we consider a one-dimensional periodic domain of the same length $L=40\pi$ as in the numerical continuation and solve Eq.~(\ref{eq:she_general}) using a semi-implicit pseudo-spectral numerical scheme for spatial derivatives \cite{uecker2009short} and a fourth-order Runge-Kutta time-stepping scheme. All DNS results presented in this paper were obtained on a finely resolved uniform spatial grid of $8192$ grid points (corresponding to approximately $100$ grid points per pattern wavelength), except when specified otherwise. We use a time step of $dt=0.01$. Larger values of $dt$ led to incorrect drift speeds.

To verify Eq.~(\ref{eq:drift_velocity}), we consider a state on the second lowest Z branch at $\epsilon=0.13$, $r=-0.66$, shown in the inset in Fig. \ref{fig:verify_velocity}. The state is asymmetric: a small change in peak/trough amplitudes with increasing $\epsilon$ can be discerned, a consequence of symmetry breaking when $\epsilon>0$.
%: at $\epsilon=0$ the state is antisymmetric about its midpoint, while it becomes asymmetric at $\epsilon>0$.

%We vary $\epsilon$ from small to large values. 
Figure \ref{fig:verify_velocity} shows the drift velocity $c$ as a function of $\epsilon$, with excellent quantitative agreement between DNSs, numerical continuation and theory for $\epsilon\lesssim 0.1$. For $\epsilon \gtrsim0.1$, the numerically obtained drift velocities deviate from the asymptotics, with the asymptotic prediction overestimating the numerical value by less than $10\%$. However, the DNSs and numerical continuation continue to show excellent agreement. At $\epsilon\approx0.13$ the state under consideration ceases to exist, since beyond this point the bifurcation structure of the system is qualitatively altered (cf.~\cite{houghton2011swift}). Additional runs with the smaller time step $dt=0.001$ were also performed, and gave the same results as those for $dt=0.01$, indicating that the DNSs are well resolved in time. We also repeated the DNSs on a coarser grid with $2048$ grid points and obtained drift velocities $c$ that are indistinguishable from those shown in Fig.~\ref{fig:verify_velocity}, indicating that the simulations are also well resolved in space. 
%%%%%%%%%%%%%%%%%%%%%%%%%%%%%%%%%%%%%%%%%%%%%%%%%%%%%%%%%%%%%%%
\begin{figure*}[hbt!]
    \centering
    \includegraphics[width=1.05\textwidth]{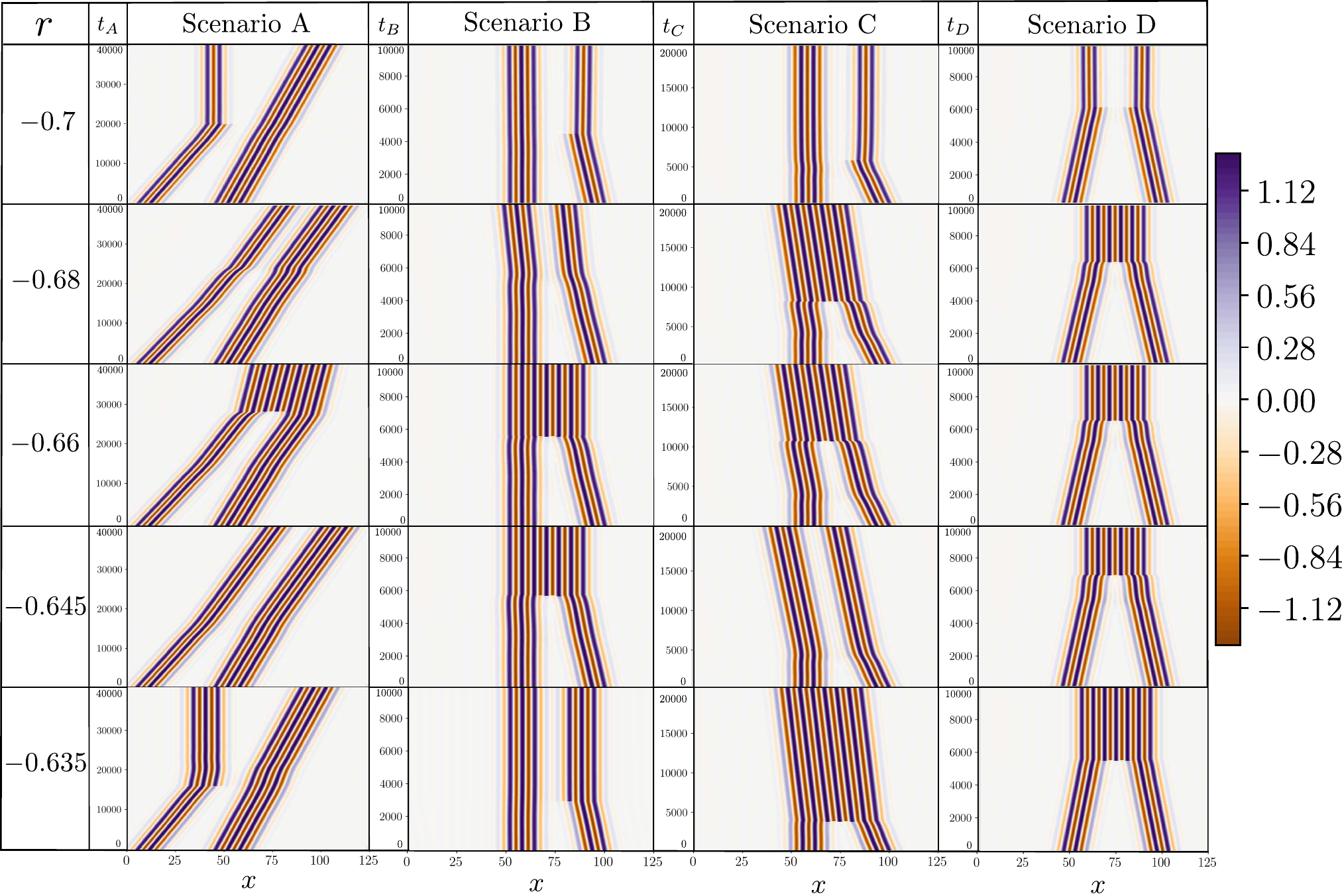}
    \caption{Space-time plots of collision scenarios A-D for five different values of $r$ indicated by gray dashed vertical lines in Fig.~\ref{fig:Delta_L2_norm}, shown with time along the vertical axis and space along the horizontal axis. The different collision scenarios involve different time scales, therefore distinct time axes are specified for each case. A rich zoology of different collision outcomes is found: one or two extrema may be deleted (for $r$ near $-0.7)$, or bound states can form without the creation or deletion of any extrema (in all scenarios except scenario D), or one, three, four or five extrema may be added in the collision, depending on the value of $r$ and on the scenario. Some of the resulting states travel while others are symmetric and hence stationary. }
    \label{fig:collision_table}
\end{figure*}
%%%%%%%%%%%%%%%%%%%%%%%%%%%%%%%%%%%%%%%%%%%%%%%%%%%%%%%%%%%%%%%

\begin{figure}[hbt]
    \centering
    \includegraphics[width=0.5\textwidth]{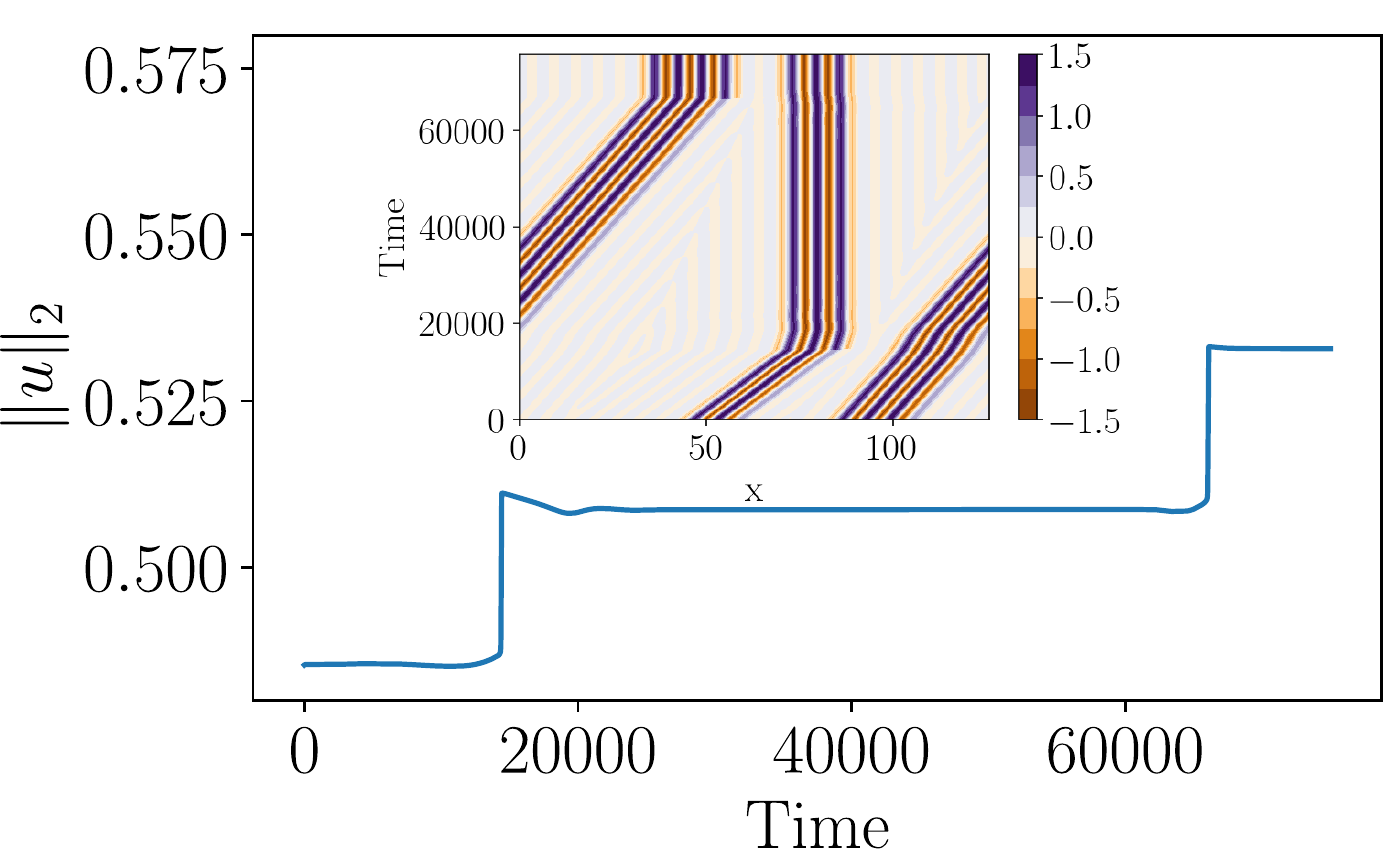}
    \caption{The $L_2$ norm versus time for a collision between two chasing asymmetric structures from scenario A ($r=-0.635$) consisting of two stages. In the first stage, the smaller asymmetric LS is rendered symmetric and hence stationary by the addition of an extremum. In the second stage, the larger asymmetric LS suffers the same fate resulting in a stationary bound state.}
    \label{fig:long_scenario_A_635}
\end{figure}

%%%%%%%%%%%%%%%%%%%%%%%%%%%%%%%%%%%%%%%%%%%%%%%%%%%%%%%%%%%%%%%

%%%%%%%%%%%%%%%%%%%%%%%%%%%%%%%%%%%%%%%%%%%%%%%%%%%%%%%%%%%%%%%

\begin{figure}[hbt]
    \centering
    \includegraphics[width=0.5\textwidth]{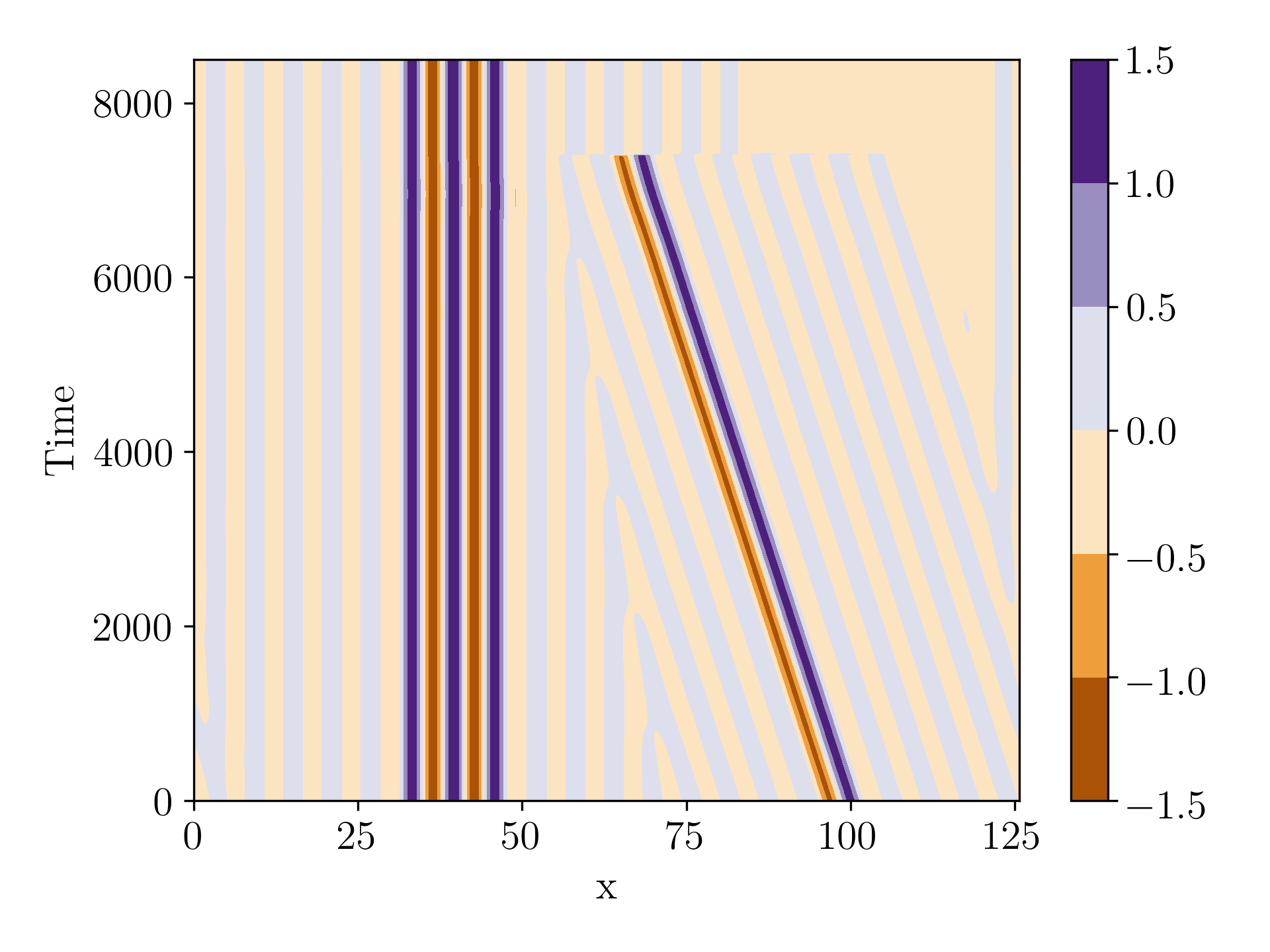}
    \caption{Annihilation of a single-wavelength asymmetric structure in collision with a symmetric LS at $r=-0.65$. This is a special case of scenario B with an asymmetric structure consisting of only two extrema, located at the leftmost edge of the range of stability of the lowest Z branch in Fig.~\ref{fig:eps_03_bif}. Note the coarser colorbar compared to Fig.~\ref{fig:collision_table}. }
    \label{fig:annihilation}
\end{figure}

%%%%%%%%%%%%%%%%%%%%%%%%%%%%%%%%%%%%%%%%%%%%%%%%%%%%%%%%%%%%%%%%%%%%%%%%%%%%%%%%%%%%%%
\section{Collisions of localized structures}
\label{sec:overview_collisions}
%%%%%%%%%%%%%%%%%%%%%%%%%%%%%%%%%%%%%%%%%%%%%%%%%%%%%%%%%%%%%%%%%%%%%%%%%%%%%%%%%%%%%%%
As shown in Fig.~\ref{fig:eps_03_bif}, multiple stable LSs can coexist in the domain, and these may undergo collisions. In this section we present the results of extensive DNSs of such collisions between different types of LSs performed using the numerical solver described in the previous section. For all results described below, we take $\epsilon=0.03$, unless stated otherwise.

%%%%%%%%%%%%%%%%%%%%%%%%%%%%%%%%%%%%%%%%%%%%%%%%%%%%%%%%%%%%%%%%%%%%%%%%%%%%%%%%%%%%%%%
\subsection{Overview of DNS results}
\label{ssec:overview}
%%%%%%%%%%%%%%%%%%%%%%%%%%%%%%%%%%%%%%%%%%%%%%%%%%%%%%%%%%%%%%%%%%%%%%%%%%%%%%%%%%%%%%%
Here we describe the rich collision phenomenology that is observed when different types of LSs collide at various values of $r$, as illustrated in Fig.~\ref{fig:collision_table}. We distinguish the following four collision scenarios:
\begin{itemize}
\item Scenario A: collision between two asymmetric states which differ in the number of significant wavelengths, and travel in the same direction, with the shorter, faster LS catching up to the longer, slower LS, as predicted by Eq.~(\ref{eq:drift_velocity}). The two colliding extrema are of opposite sign. We focus specifically on collisions between two LSs of length two and three wavelengths. Four different possible outcomes are observed: deletion of one extremum, formation of a drifting bound state without a change in the number of extrema, and the creation of one or four new extrema. Note that the bound states at $r=-0.68$ and $r=-0.645$ shown in Fig.~\ref{fig:collision_table} differ in their separation. The cases $r=-0.7$ and $r=-0.635$ in fact involve two \textit{separate} consecutive collisions due to the periodicity of the domain, of which only the first is shown in Fig.~\ref{fig:collision_table}. In the former case, the second collision (not shown) results in a drifting bound state, while in the latter case (shown in Fig.~\ref{fig:long_scenario_A_635}), the asymmetric LS is rendered symmetric and stationary owing to the nucleation of an additional extremum, resulting in a larger but stationary bound state.

\item Scenario B: a drifting asymmetric state collides with a stationary symmetric state with a maximum at its center. The two colliding extrema are of opposite sign. We focus on a two-wavelength asymmetric LS and a symmetric LS with three positive and two negative large-amplitude extrema. Four different possible collision outcomes are observed: deletion of one extremum, formation of a drifting bound state without change in the number of extrema, and the creation of one or four new extrema. 

\item Scenario C: same as scenario B but with the symmetric LS flipped by $u\to-u$ so that the two colliding extrema are of the same sign. We focus on a two-wavelength asymmetric LS and a symmetric structure with three negative and two positive large-amplitude extrema. Five different possible collision outcomes are observed: deletion of one extremum, formation of a drifting bound state without change in the number of extrema, and the creation of one, three or five new extrema. 

\item Scenario D: head-on collision between two asymmetric states. The two colliding extrema are of the same sign. We focus on a pair of identical two-wavelength patterns (collisions between asymmetric LSs of distinct sizes were also considered, and gave qualitatively similar results). Four different possible collision outcomes were observed: deletion of two extrema and the creation of three or five new extrema. No pure bound states were observed in this case. 
\end{itemize}

Figure \ref{fig:collision_table} shows a summary of the space-time plots depicting the different possible collision outcomes. All collisions were simulated using DNS on a uniform grid of $2048$ points to facilitate longer simulation times. Some cases were also repeated with $8192$ grid points and no change in the collision behavior was observed.

Three qualitatively distinct collision outcomes are observed across all collision scenarios: deletion of extrema, formation of a bound state, and the creation of new extrema. Some of these states travel while others are symmetric and hence stationary. One notices that scenarios A and B are quite similar to one another in terms of the outcome realized at a given $r$ (except near $r=-0.645$, see also Fig.~\ref{fig:Delta_L2_norm}). Scenarios C and D are similar in the same sense. Since the closest extrema in the collisions in scenarios A and B are both of opposite sign but are of the same (negative) sign in scenarios C and D, one might surmise that the relative sign of interacting extrema is important in determining the qualitative collision outcome, cf.~\cite{houghton2011swift}. Our detailed results broadly support this suggestion but indicate that this sign is not the sole factor determining the collision outcome since differences between scenarios persist.

%The size of colliding LSs may have a qualitative impact on the collision outcome. For example, Fig.~\ref{fig:annihilation} shows a special case of scenario B, where a single-wavelength asymmetric LS collides with a stationary symmetric structure and is annihilated. This is because this asymmetric LS is located on the lowest Z branch in Fig.~\ref{fig:eps_03_bif} and so is minimal in the sense that no stable LS exists with less than one significant wavelength. Hence, the deletion of an extremum implies annihilation of this LS. Another example is found in scenario $A$ at $r=-0.7$, where a traveling bound state arises after the second collision (not shown) between a three-wavelength asymmetric LS and a symmetric LS with a single minimum. Such bound states are not observed in any other case at $r=-0.7$ (cf. Fig.~\ref{fig:collision_table}). We leave the systematic investigation of the dependence of collision outcomes on the size of colliding LSs to a future study.

Figure~\ref{fig:collision_table} reveals that the speed $c$ of the traveling state plays a significant role in determining the collision outcome. This speed is controlled by the value of the parameter $r$ since $r$ controls the degree of asymmetry of each traveling state, but it also depends on the length of the structure, cf.~Fig.~\ref{fig:c_vs_r}. For example, in scenario A at $r=-0.7$ a narrow LS catches up to a wider LS, with the resulting interaction stopping the former and leaving the latter unaffected. Because of periodic boundary conditions the wider, slower moving LS collides with the stationary state in a subsequent collision, creating a drifting bound state (not shown, but see Fig.~\ref{fig:long_scenario_A_635} for a similar multiple collision at $r=-0.635$). Such drifting bound states are not observed in any of the other scenarios at this value of $r$ where a narrower and so faster LS interacts with a broader LS at rest. Figure~\ref{fig:annihilation} shows an extreme case of scenario B, where a single-wavelength asymmetric LS collides with a stationary symmetric structure and is annihilated. This is because this asymmetric LS is located on the lowest Z branch in Fig.~\ref{fig:eps_03_bif} and so is minimal in the sense that no stable LS shorter than one significant wavelength exists. Hence, the deletion of an extremum implies annihilation of this LS. 

%%%%%%%%%%%%%%%%%%%%%%%%%%%%%%%%%%%%%%%%%%%%%%%%%%%%%%%%%%%%%%%%%%%%%%%%%%%%%%%%%%%%%%%
\subsection{Change in $L_2$ norm before and after collision}
\label{ssec:delta_norm}
\begin{figure}
    \centering
    \includegraphics[width=0.5\textwidth]{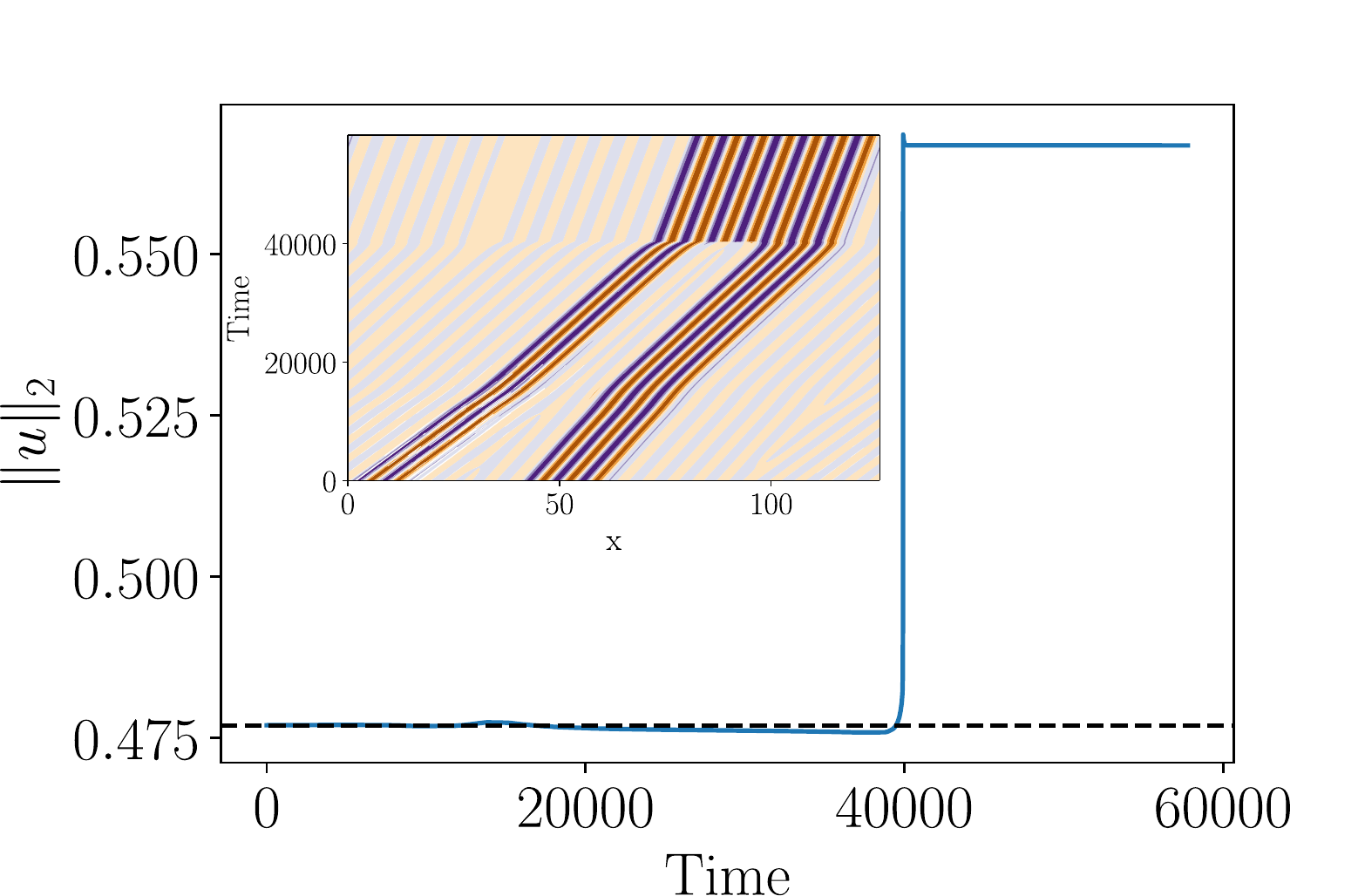}
    \caption{{The $L_2$ norm versus time for a collision of two chasing asymmetric LSs (scenario A) at $r=-0.65$. The collision of the two chasing structures occurs at $t\approx20~000$, leading to the creation of a traveling metastable bound state, whose structure changes slowly over time, as visible from the slope in $\|u\|_2$. At $t\approx 40~000$,  three additional extrema are nucleated by nonlinear interactions, behavior resulting in the jump in $\|u\|_2$.}}
    \label{fig:L2_norm_vs_time}
\end{figure}

\begin{figure}
    \centering
    \includegraphics[width=0.5\textwidth]{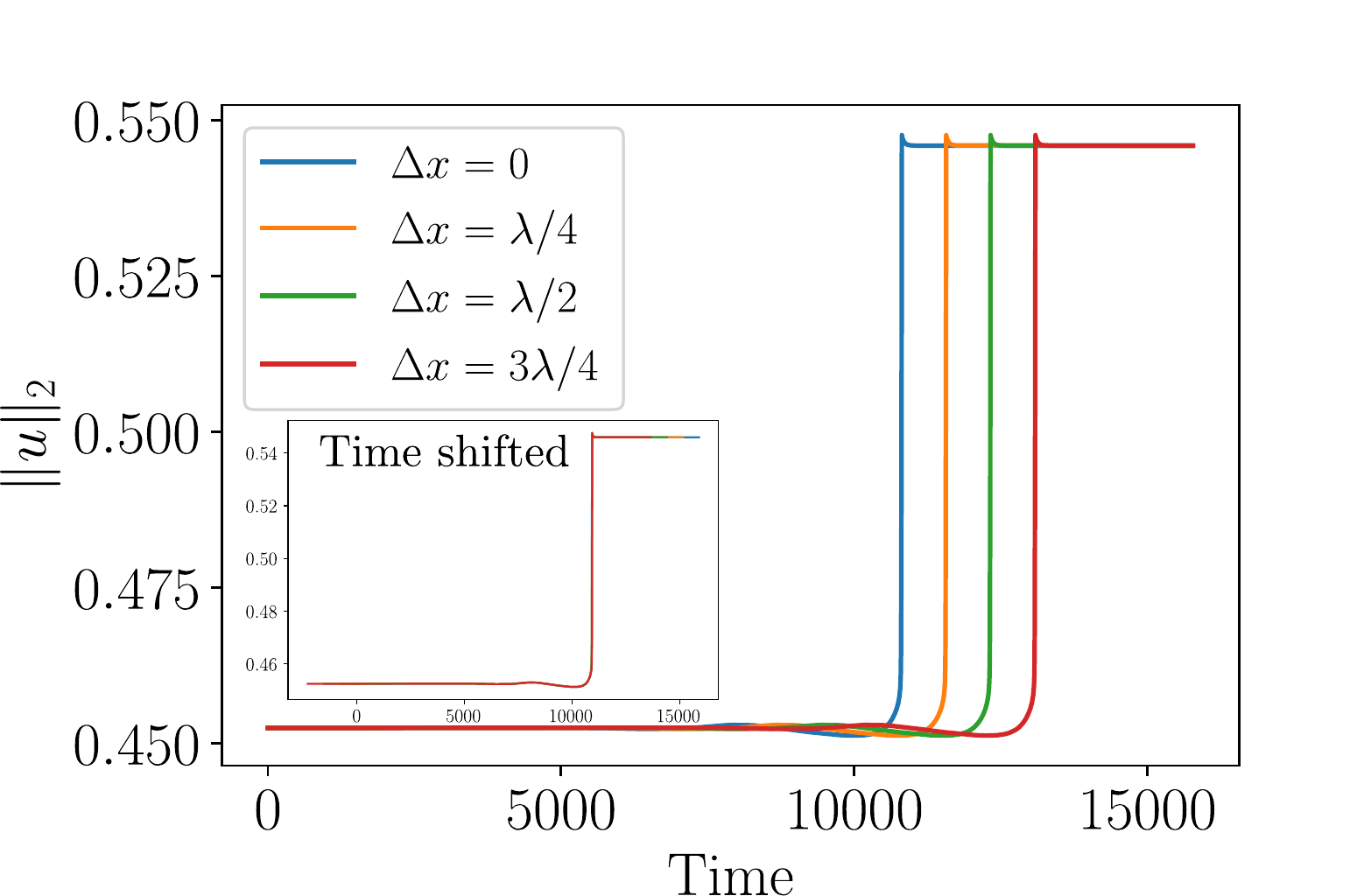}
    \caption{The $L_2$ norm versus time for symmetric-asymmetric collisions from scenario B (at $r=-0.65)$ for initial conditions shifted by different fractions $\Delta x$ of the wavelength $\lambda=2\pi/\beta$, with $\beta$ the imaginary part of the spatial eigenvalue defined in Eq.~(\ref{eq:spatial_eigenvalues}). Inset shows the same data with time shifted by $\Delta t = \Delta x/c$, where $c$ is the drift speed of the asymmetric LS: all curves collapse exactly, indicating that the collision dynamics are independent of initial separation.}
    \label{fig:varying_initial_distance}
\end{figure}

%%%%%%%%%%%%%%%%%%%%%%%%%%%%%%%%%%%%%%%%%%%%%%%%%%%%%%%%%%%%%%%%%%%%%%%%%%%%%%%%%%%%%%%
To quantify the change in pattern size during a collision, we measure the $L_2$ norm defined in Eq.~(\ref{eq:def_L2}). Figure~\ref{fig:L2_norm_vs_time} shows a sample time series of $\|u\|_2$ from a chasing collision (scenario A) at $r=-0.65$, where the collision first produces a metastable bound state, but three new extrema are generated at late times by nonlinear interactions. Figure~\ref{fig:L2_norm_vs_time} suggests the simple metric
\begin{equation}
    \Delta \|u\|_2 = \|u_{\textrm{final}}\|_2 - \|u_{\textrm{initial}}\|_2,
    \label{eq:delta_norm}
\end{equation}
where $u_{\textrm{initial}}$ is the state before the collision, i.e., when the distance between the two structures is still large, and $u_{\textrm{final}}$ is the resulting state a long time after the collision has occurred. Importantly, the collision dynamics are independent of the initial distance between the colliding LSs, since there is no inertia in the system. This is illustrated in Fig.~\ref{fig:varying_initial_distance}, where the initial distance in scenario B (symmetric-asymmetric collision) at $r=-0.65$ is varied by fractions of the wavelength associated with the spatial eigenvalue, $\lambda=2\pi/\beta$, with $\beta$ defined in Eq.~(\ref{eq:spatial_eigenvalues}). Figure~\ref{fig:varying_initial_distance} shows that the collision dynamics are self-similar under shifts in the initial distance: simply accounting for the time required to propagate over the additional distance leads to data collapse. This has also been verified explicitly for other collisions, including from scenario D. We deduce from these observations that the only variables affecting collision outcomes are indeed the chosen pair of colliding structures, and the control parameter $r$ (at fixed $\epsilon$).

The phenomenology illustrated in Fig.~\ref{fig:L2_norm_vs_time} is reminiscent of collision dynamics described in terms of \textit{scattors} \cite{nishiura2003dynamic,nishiura2005scattering}: unstable stationary or time-periodic patterns which direct the evolution in state space during the collision process along their stable and unstable manifolds. While these ideas were proposed in the context of models other than SH35, they may be applicable for some of the collision events observed here.

%%%%%%%%%%%%%%%%%%%%%%%%%%%%%%%%%%%%%%%%%%%%%%%%%%%%%%%%%%%%%%%
%%%%%%%%%%%%%%%%%%%%%%%%%%%%%%%%%%%%%%%%%%%%%%%%%%%%%%%%%%%%%%%
%%%%%%%%%%%%%%%%%%%%%%%%%%%%%%%%%%%%%%%%%%%%%%%%%%%%%%%%%%%%%%%
%%%%%%%%%%%%%%%%%%%%%%%%%%%%%%%%%%%%%%%%%%%%%%%%%%%%%%%%%%%%%%%

\begin{figure}[h]
     \centering
     \includegraphics[width =0.5\textwidth]{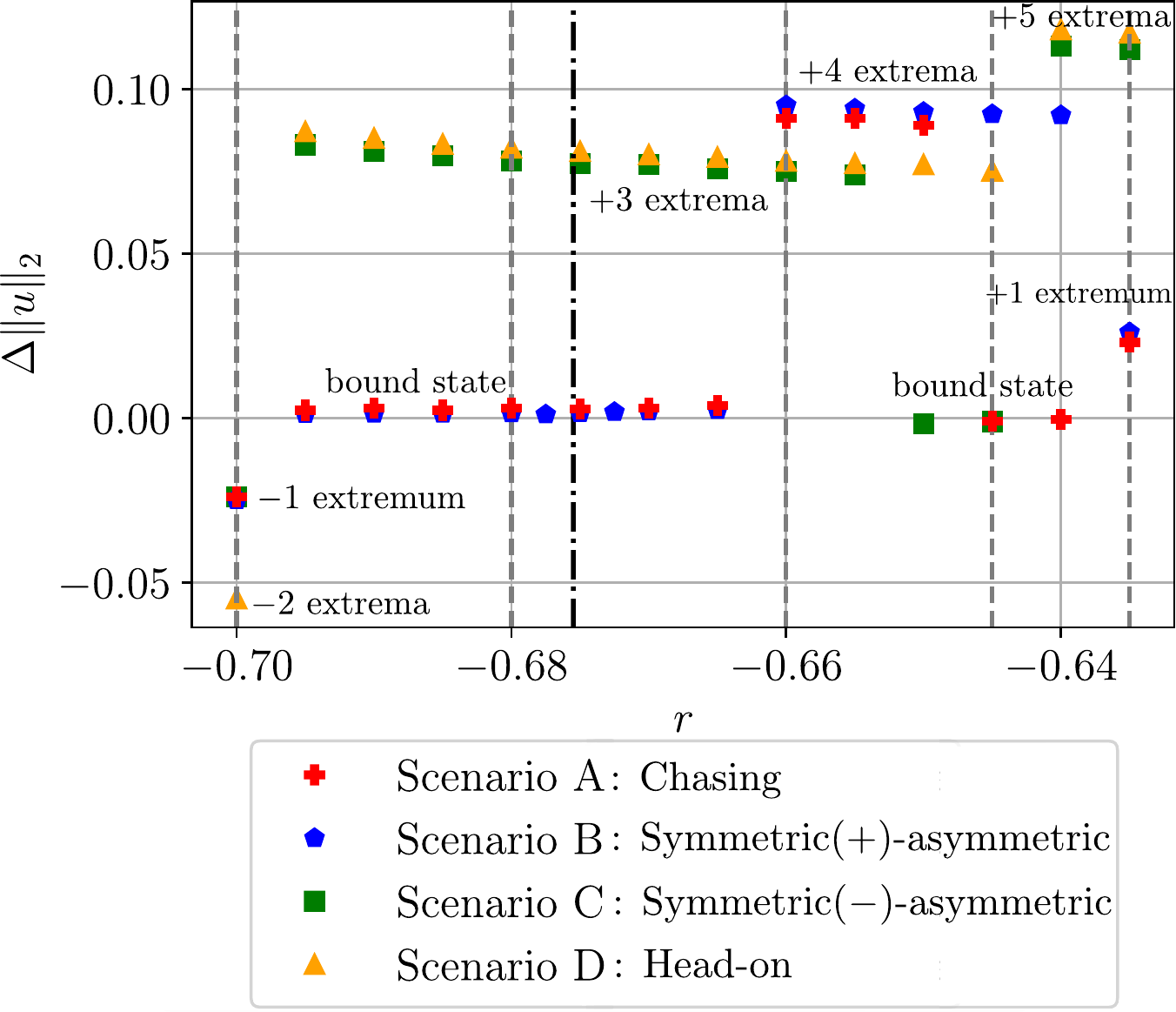}
     \caption{Change in the $L_2$ norm before and after a collision, $\Delta \|u\|_2$ defined in Eq.~(\ref{eq:delta_norm}), for different collision scenarios, shown as a function of $r$. Labels indicate the number of extrema gained or lost in the collision. Vertical dashed gray lines correspond to the values of $r$ shown in Fig.~\ref{fig:collision_table}. The vertical black dash-dotted line shows the location of the Maxwell point $r_M$ for $\epsilon=0$ (all remaining results shown are for $\epsilon=0.03$).}
     \label{fig:Delta_L2_norm}
 \end{figure}

Figure~\ref{fig:Delta_L2_norm} shows $\Delta \| u\|_2$ as a function of $r$ for all four scenarios. A value of $\Delta\|u\|_2\approx0$ corresponds to the formation of a bound state without a change in the overall pattern size, a positive value indicates the creation of one or more extrema, while a negative value indicates that one or more extrema were deleted.

The range of $r$ shown in Fig.~\ref{fig:Delta_L2_norm} corresponds to the interval with stable propagating solutions when $\epsilon=0.03$. Vertical gray dashed lines indicate the cases depicted in Fig.~\ref{fig:collision_table}.  Near the leftmost edge of the existence interval, at $r=-0.7$, one (A-C) or two extrema (D) are deleted in the collision. Near the rightmost edge, at $r=-0.635$, new extrema are created: either one (A, B) or five extrema (C, D). Away from these boundary cases, the scenarios differ more substantially. At $r=-0.68$, one observes either the formation of bound states (A, B), or the creation of three additional extrema (C, D). At $r=-0.66$, new extrema are added in the collision in all cases: either four additional extrema (A, B) or three (C, D). At $r = -0.645$ and $r=-0.65$ (not shown), scenarios A and C revert to $\Delta \| u \|_2\approx 0$, i.e. the formation of a bound state, while new extrema continue to be added in scenarios B (four extrema) and D (three extrema).

As already mentioned in the discussion of Fig.~\ref{fig:collision_table}, scenarios A and B are similar to one another, as are C and D, in the sense that for most values of $r$, they display the same number of extrema added or deleted. However, Fig.~\ref{fig:Delta_L2_norm} reveals deviations between these scenarios in the interval $-0.65\lesssim r \lesssim -0.64$. To ensure that these are not numerical artefacts, we repeated the runs in scenarios A and B in this range of $r$ at higher spatial resolution, using $8192$ instead of $2048$ grid points, and continued the simulation until very late times ($t\approx 150~000$). We also repeated the runs with significantly smaller time steps, $dt=0.005$ and $dt=0.001$, with the same collision outcome, confirming that the nonmonotonic dependence of $\Delta \| u\|_2$ on $r$ is a robust result that we discuss further below.

The black dash-dotted vertical line in Fig.~\ref{fig:Delta_L2_norm} indicates the location of the Maxwell point $r=r_M$ for the variational case $\epsilon=0$. The figure shows that the values of $r$ where $\Delta\| u \|_2$ changes differ between scenarios and that, in addition, these locations lie far from $r=r_M$. We conclude that the Maxwell point of the variational case has little, if any, relevance in determining the collision outcome in the nonvariational case, even in the weak symmetry-breaking case considered here. 

%We remind the reader that the free energy $\mathcal{F}$ of the stable periodic pattern state decreases monotonically with $r$ as shown in Fig.~\ref{fig:free_energy}. 
The overall trend of increasing $\Delta \| u\|_2$ with increasing $r$ can be viewed as a reflection, in the nonvariational regime, of similar behavior in the variational case: when $\epsilon=0$, the free energy $\mathcal{F}$ of the stable pattern state decreases with increasing $r$, as shown in Fig.~\ref{fig:free_energy}, and the periodic pattern becomes increasingly energetically favored. Apart from some exceptions at $r\geq -0.65$ (to be discussed further below), the increasing trend in the number of extrema added in a collision is compatible with this intuition. Finally, even for a fixed number of extrema added or lost, Fig.~\ref{fig:Delta_L2_norm} reveals a slow decrease in $\Delta \| u \|_2$ with increasing $r$. This is associated with small changes in the profile of the extrema as $r$ changes. 
 %%%%%%%%%%%%%%%%%%%%%%%%%%%%%%%%%%%%%%%%%%%%%%%%%%%%%%%%%%%%%%%
 %%%%%%%%%%%%%%%%%%%%%%%%%%%%%%%%%%%%%%%%%%%%%%%%%%%%%%%%%%%%%%%

%%%%%%%%%%%%%%%%%%%%%%%%%%%%%%%%%%%%%%%%%%%%%%%%%%%%%%%%%%%%%%%%%%%%%%%%%%%%%%%%%%%%%%%
\subsection{Bound states: isolas and stability}
\label{ssec:isolas}
%%%%%%%%%%%%%%%%%%%%%%%%%%%%%%%%%%%%%%%%%%%%%%%%%%%%%%%%%%%%%%%%%%%%%%%%%%%%%%%%%%%%%%%

Figure-eight isola structures extending over the entire width of the snaking region have been previously described in the quadratic-cubic Swift-Hohenberg equation without nonvariational terms, arising from multi-pulse solutions consisting of two LSs bound together in close proximity \cite{burke2009multipulse,beck2009snakes}. Here, we find that for bound states formed from collisions at $\epsilon>0$, the properties of the isolas to which these states belong depend strongly on the collision scenario.

\begin{figure}
    \centering
    \includegraphics[width = 0.5\textwidth]{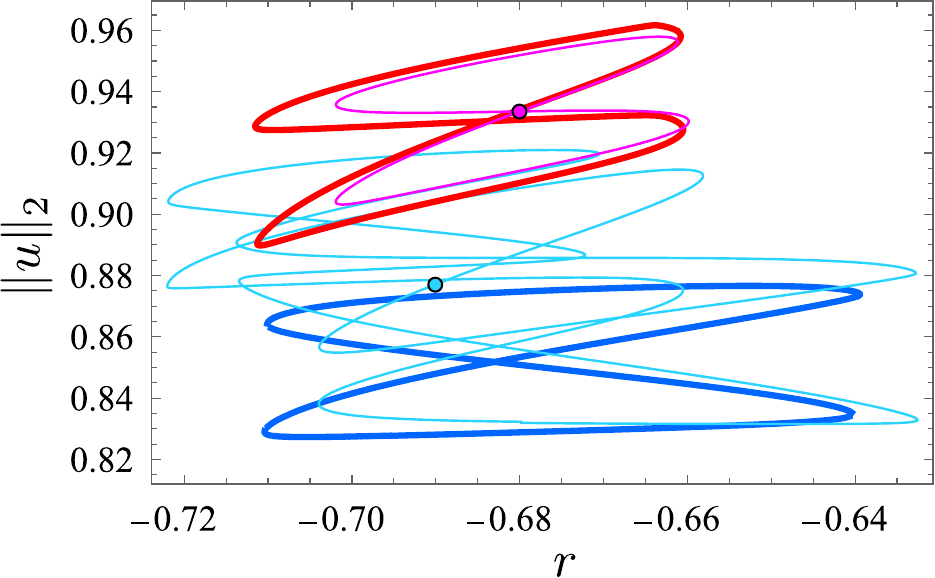}
    \caption{Traveling bound states obtained from numerical continuation of collision results in scenarios A and B at $\epsilon=0.03$. Thin pink line: traveling state obtained from a chasing collision (scenario A) at $r=-0.68$. Thin light blue line: similar traveling state obtained from a symmetric-asymmetric collision (scenario B) at $r=-0.69$. Starting points for continuation in $r$ are marked with circles.  Continuation in $\epsilon$ to the variational case $\epsilon=0$ yields isolas of stationary two-pulse states. Thick red line: scenario A at $\epsilon=0$. Thick dark blue line: scenario B at $\epsilon=0$.}
    \label{fig:isolas}
\end{figure}
%%%%%%%%%%%%%%%%%%%%%%%%%%%%%%%%%%%%%%%%%%%%%%%%%%%
%%%%%%%%%%%%%%%%%%%%%%%%%%%%%%%%%%%%%%%%%%%%%%%%%%%
\begin{figure}[hbt]
    \centering
    \includegraphics[width=0.5\textwidth]{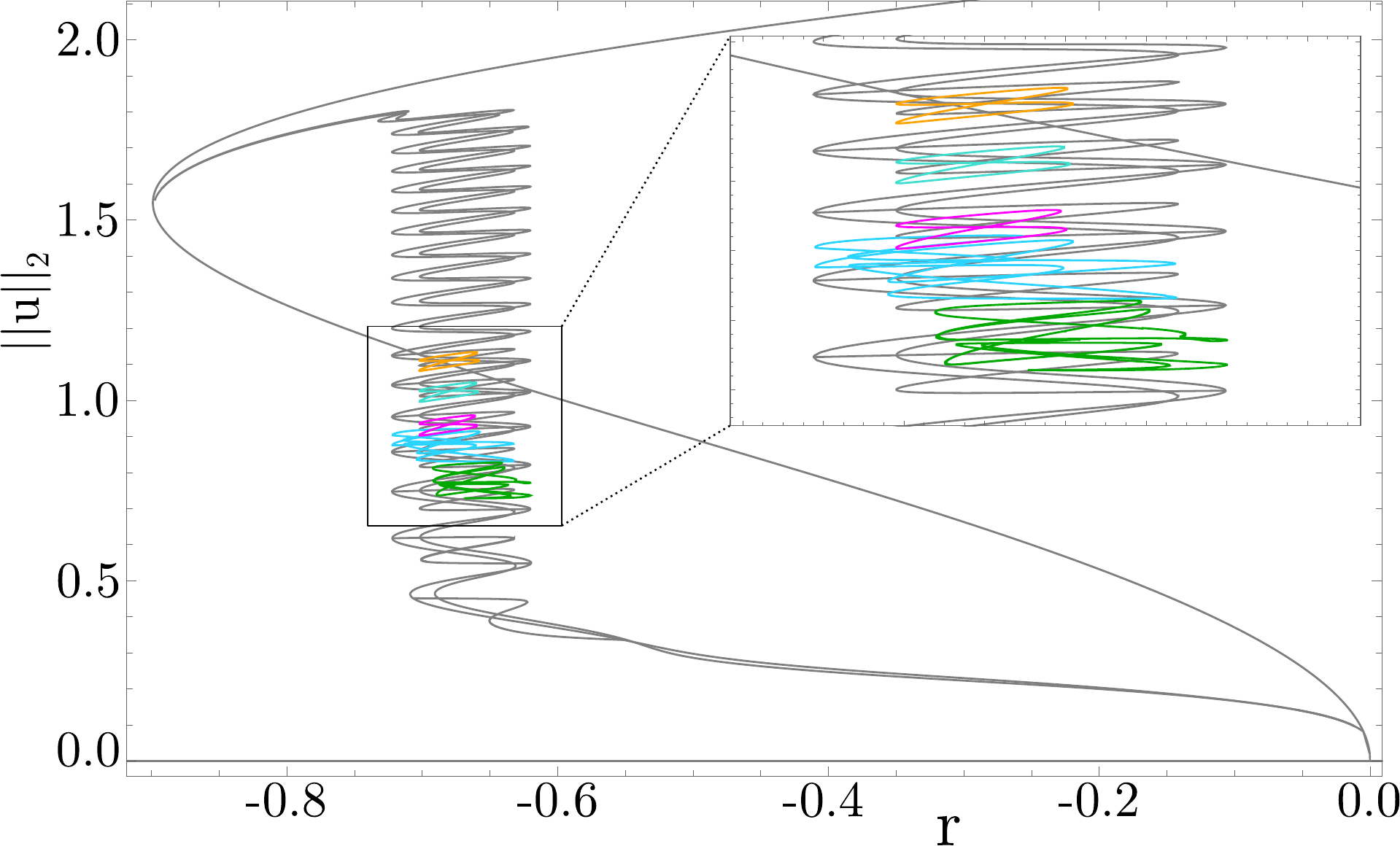}
    \caption{Isola structures (colored) corresponding to the traveling two-pulse bound states generated in different collision scenarios superposed on the bifurcation diagram from Fig.~\ref{fig:eps_03_bif} (gray). All results are for $\epsilon=0.03$. }
    \label{fig:isola_full_BD}
\end{figure}
%%%%%%%%%%%%%%%%%%%%%%%%%%%%%%%%%%%%%%%%%%%%%%%%%%%
Figure~\ref{fig:isolas} shows two isolas at $\epsilon=0.03$ (thin lines) obtained from a continuation in $r$, starting from two bound states generated by a collision in scenarios A and B at $r=-0.68$ and $r=-0.69$, respectively (points highlighted in Fig.~\ref{fig:isolas}), together with the results of continuing these isolas in $\epsilon$ to the variational case $\epsilon=0$ (thick lines). The figure shows that in the former case (red curves) the isola retains its shape as it transforms into the corresponding isola of the variational problem. In the latter and more typical case, the isola at $\epsilon=0.03$ takes the form of a spaghetti-like tangle (light blue), requiring substantial simplification prior to reaching the corresponding isola of the variational problem (thick blue). Figure~\ref{fig:isola_full_BD} shows several additional isolas of traveling bound states at $\epsilon=0.03$ obtained by colliding longer initial LSs and superposed on the full bifurcation diagram (colors match between Figs.~\ref{fig:isolas} and \ref{fig:isola_full_BD}). All figure-eight isolas seen in Fig.~\ref{fig:isola_full_BD} are from scenario A, while tangled isolas are from scenario B.
%It is seen that the shape of the isolas to which the bound states produced in different collision scenarios belong differs qualitatively between scenarios at $\epsilon>0$. Specifically, in scenario A the figure-eight isola from $\epsilon=0$ is retained, while in other cases the isolas are significantly more complex at $\epsilon>0$, forming a spaghetti-like tangle. 

\begin{figure}[hbt]
    \centering
    \includegraphics[width=0.5\textwidth]{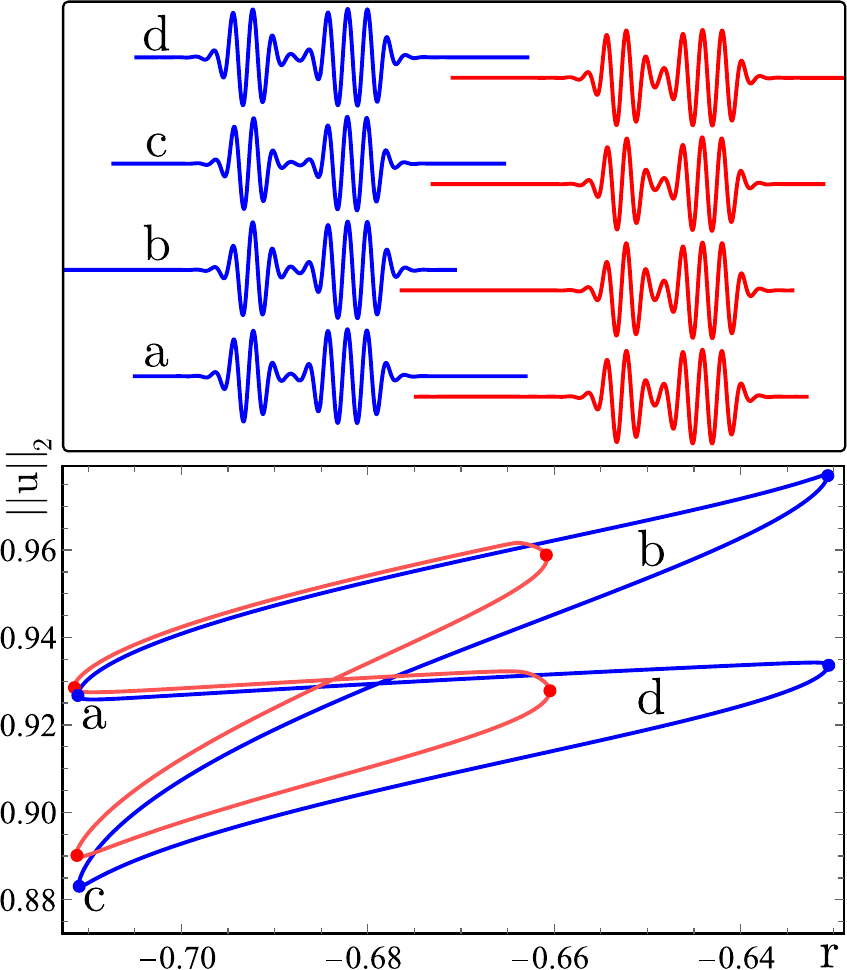}
    \caption{The $L_2$ norm vs. control parameter $r$ for two isolas at $\epsilon=0$, obtained by numerical continuation from traveling bound states in scenario A with $\epsilon=0.03$ resulting from collisions at $r=-0.645$ (large isola) and $r=-0.68$ (small isola), respectively. Color-matched solution profiles are shown in the top panel, where the red solutions (right), corresponding to the small isola, feature a separation between the two single-pulse LSs shorter by one wavelength as compared with the blue profiles (left), which correspond to the large isola. }
    \label{fig:two_isolas}
\end{figure}

It is important to note that the rightmost edge of the figure-eight isola in scenario A shown in Fig.~\ref{fig:isolas} (thin red line) is located near $r\approx-0.66$, and not at $r\approx-0.63$, the right boundary of the snaking region for $\epsilon=0.03$. This is a consequence of the fact that two-pulse states with a small separation between the LSs lie on isolas that are significantly smaller than those for bound states with a larger separation. This is so not only when $\epsilon=0$ \cite{burke2009multipulse} but also when $\epsilon=0.03$. Figure~\ref{fig:two_isolas} shows an example of a narrow figure-eight isola at $\epsilon=0$ obtained from a bound state at $\epsilon=0.03$ and $r=-0.68$ together with a wider figure-eight isola obtained by continuing a bound state, also obtained from a collision in scenario A with $\epsilon=0.03$, but at $r=-0.645$. Despite their similarity (see profiles in Fig.~\ref{fig:two_isolas}, top panel) these profiles are part of separate isolas: the solution profiles of the narrow isola feature an interaction region between the two LSs comprising the bound state which is one wavelength shorter than that of the larger isola solution profiles, a property that is preserved upon numerical continuation. As the separation between the two LSs decreases and their interaction becomes stronger, the figure-eight isolas shrink, with the rightmost edge moving farther and farther to the left.

%This decrease is size plays a key role in determining collision outcomes: as shown in Fig.~\ref{fig:isolas} and discussed above, the rightmost edge of the figure-eight isola coincides with the value of $r$ where collision outcomes transition from pure bound states to the addition of new extrema in scenario A.

This decrease in size plays a key role in determining collision outcomes: the value $r\approx-0.66$ (Fig.~\ref{fig:isolas}, thin read line) coincides with the parameter value in Fig.~\ref{fig:Delta_L2_norm} where new extrema are first created in scenario A. This observation suggests the following way of understanding the observed phenomenology: when two LSs approach one another in a collision, there are two possibilities. Either (a) a stable multipulse bound state exists at the value of $r$ in question for the given pair of LSs, or (b) no such state exists or at least it is not stable. In case (a), the collision will lead to the formation of a stable bound state. In case (b), on the other hand, no such state is available, and the system creates or deletes extrema to reach another stable solution. 

With this hypothesis, the nonmonotonic behavior in the number of extrema created summarized in Fig.~\ref{fig:Delta_L2_norm} can be explained as follows: at parameter values $-0.65\lesssim r\lesssim -0.64$, there is an island of stability where stable bound states exist. When no stable bound state exists, a strongly nonlinear interaction must occur, resulting in LSs with a different number of extrema; this number increases with increasing $r$, as discussed earlier. 
%there is a noticeable trend for the number of peaks created in a collision (or alternatively $\|u\|_2$ to increase with $r$. This in turn can be understood as a surviving manifestation of the variational SH35 in the nonvariational regime. The free energy $\mathcal{F}$ of the global pattern state decreases monotonically with $r$ as shown in Fig.~\ref{fig:free_energy}. Therefore, excluding the cases where stable bound states are created and the boundary case $r=-0.635$, a monotonically increasing trend in the number of created peaks with $r$ is observed.

To test the hypothesis that the existence and stability of a bound state is the determining factor in setting the collision outcome, we perform stability experiments. Specifically, to verify that the bound state either does not exist or is unstable in scenarios A, B with $-0.66 \lesssim r \lesssim 0.65$ (where new extrema form), we performed DNS at these values of $r$, initializing with the post-collision final states from $r=-0.68$, $r=-0.67$, $r=-0.65$ and $r=-0.645$. In each of these cases we observed the eventual generation of additional extrema. Only one exception was observed, at $r=-0.66$ in scenario B, when a bound state from $r=-0.68$ was used as the initial condition and found to remain a pure bound state indefinitely, without any change in the number of extrema. Thus collisions may trigger the creation of new extrema despite the existence of a stable bound state, provided the bound state has only a small basin of attraction.
% {\bf especially so? In fact there are many bound states that differ by half wavelength in separation. Are you saying all of these have similar stability properties? }

We repeated the above study for scenario C: stable bound states formed from collisions at $r=-0.645$ and $r=-0.65$ were used as initial conditions at $r=-0.635$, $r=-0.64$, $r=-0.655$, $r=-0.66$. In all cases but one, the same number of extrema was spontaneously created as observed in the collision experiments. The exception, where a bound state remains indefinitely stable, was again found near the transition from a bound state to the creation of new extrema in the collision, namely at $r=-0.64$. In summary, these stability experiments largely confirm the proposed hypothesis for explaining Fig.~\ref{fig:Delta_L2_norm}: collisions yield bound states when these exist as stable states, and otherwise lead to a change in the number of extrema (typically creation of new extrema). The only deviations from this paradigm are observed when stable bound states exist but have a small basin of attraction. Then the finite perturbation provided by a collision can trigger a change in the number of extrema.

%Figure~\ref{fig:isola_full_BD} shows the isolas at $\epsilon=0.03$ from  Fig.~\ref{fig:isolas}, together with several other examples of traveling bound states obtained by colliding longer initial LSs, superposed on the full bifurcation diagram (colors match between Figs.~\ref{fig:isolas} and \ref{fig:isola_full_BD}). All figure-eight isolas seen in Fig.~\ref{fig:isola_full_BD} are from scenario A, while tangled isolas are from scenario B.
%%%%%%%%%%%%%%%%%%%%%%%%%%%%%%%%%%%%%%%%%%%%%%%%%%%

\begin{figure}
    \centering
    \includegraphics[width=\linewidth]{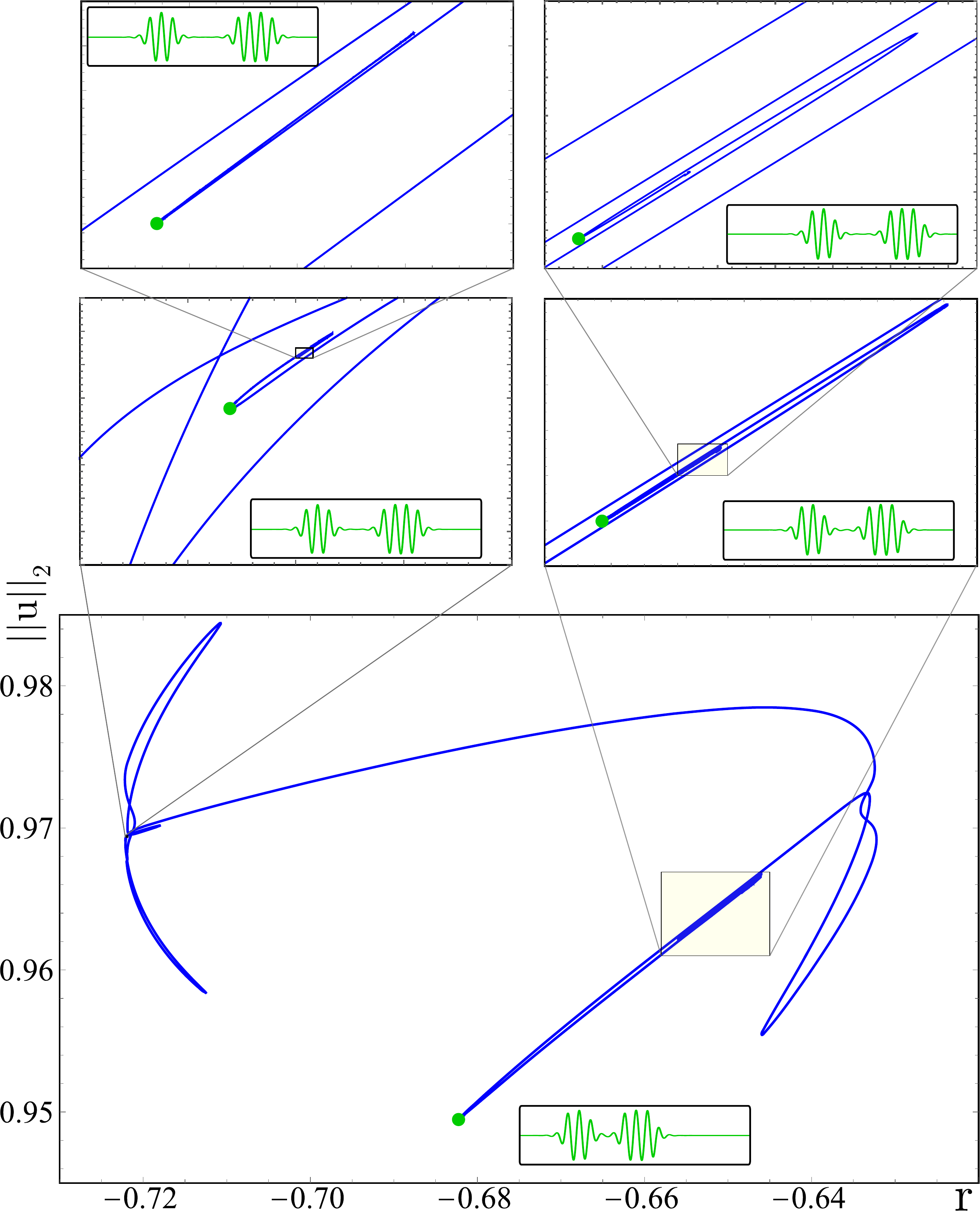}
    \caption{Scenario A bound state isola continued from $r=-0.645$ and ending in T-points at either end.  Solution profiles and parts of the spiral branch are shown at increasing levels of magnification in the upper panels.}
    \label{fig:Tpoint}
\end{figure}

\begin{figure}
    \centering
    \includegraphics[width=\linewidth]{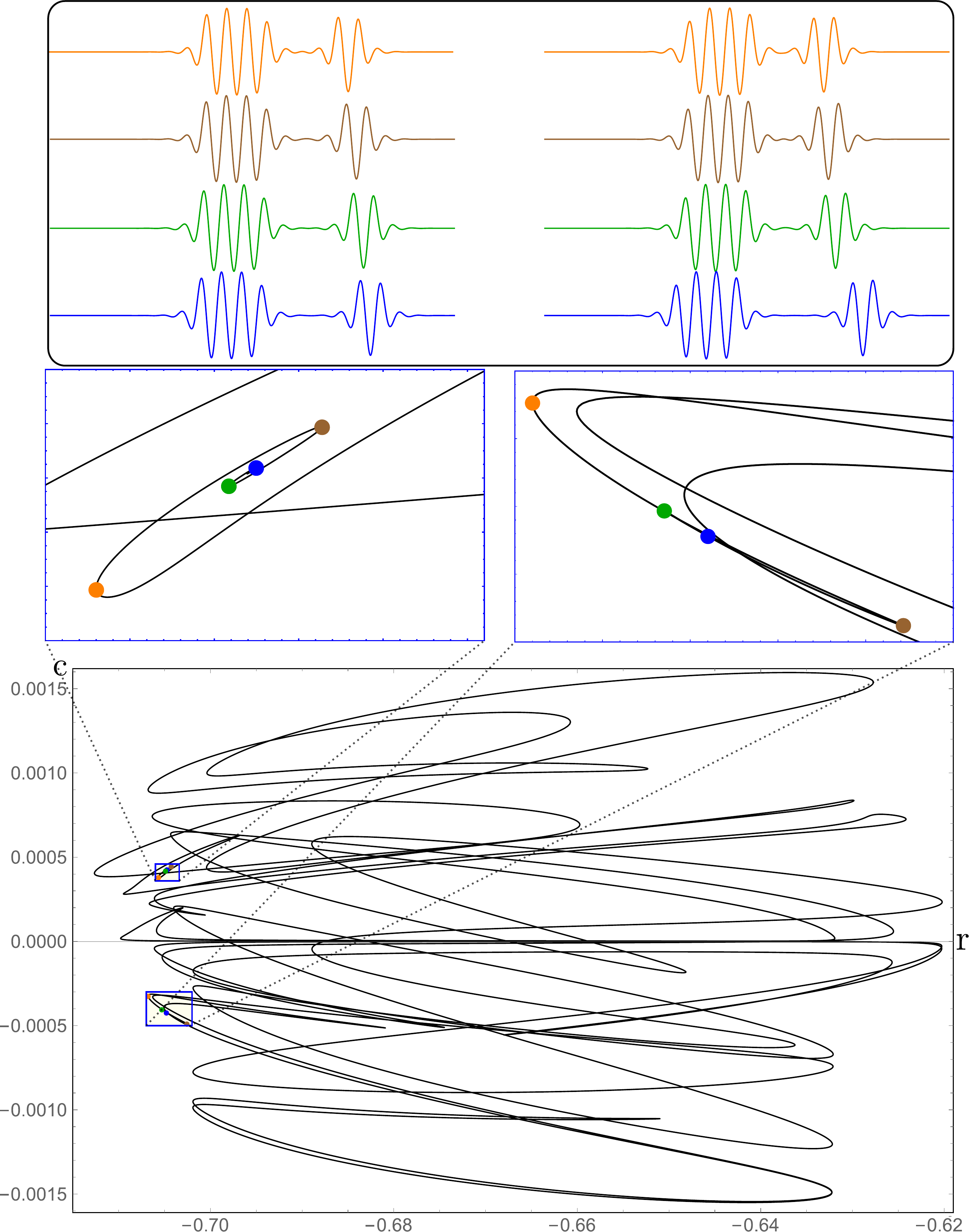}
    \caption{Scenario A bound state isola continued from $r=-0.7$ and ending in T-points at either end.  Solution profiles and parts of the spiral branch are shown at increasing levels of magnification in the upper panels.}
    \label{fig:Tpoint2}
\end{figure}

The isolas corresponding to multi-pulse bound states at $\epsilon>0$ can take other forms as well. Figures~\ref{fig:Tpoint} and \ref{fig:Tpoint2} show two such cases, in which a complex tangled branch terminates at either end in bifurcation points located at the center of a {\it spiral} structure. These figures correspond to the continuation of bound states resulting from scenario A collisions at $r=-0.645$ and $r=-0.7$, respectively. As mentioned in Sec.~\ref{ssec:overview}, the latter is in fact the result of two {\it separate} collisions. The first collision occurs between two moving LSs and leaves one stationary and symmetric LS  and one moving LS as shown in Fig.~\ref{fig:collision_table}.  The second collision (not shown) occurs after this, when the moving LS collides with the stationary LS from the other side due to periodic boundary conditions, forming a traveling bound state.
%Figure \ref{fig:Tpoint2} shows another, perhaps clearer, example of an isola which is born from and terminates in similar bifurcation points with associated spiral structures.

The bifurcations identified in Figs.~\ref{fig:Tpoint} and \ref{fig:Tpoint2} are of codimension two and occur as the separation between the two LSs in the bound state grows without limit, while the structures independently change their amplitude and degree of asymmetry such that their velocities remain the same. As the branch spirals towards the center, the separation between the constituent LSs grows. These bifurcations are highly reminiscent of T-point bifurcations, also known as Bykov points \cite{bykov1980bifurcations,bykov1993bifurcations}, which have been described as corresponding to an equilibrium-to-equilibrium heteroclinic cycle, with two branches of primary homoclinic orbits bifurcating from it. Bifurcations of this type have been observed in a variety of systems, including the Lorenz63 system \cite{glendinning1986t}, a model of calcium pulses in pancreatic cells \cite{ROMEO2003242}, and in type-I excitable media \cite{moreno2022bifurcation}. In contrast, here the two constituent LSs individually approximate a homoclinic orbit to the trivial state $u=0$ and at the T-point the state of the system corresponds to a double homoclinic cycle. Of course, in a finite domain we cannot reach this T-point, and our numerical continuation results therefore only produce a periodic structure that approximates this double homoclinic cycle. To the best of our knowledge, a T-point bifurcation from such a double homoclinic cycle described here has not been observed previously. Note that such T-points cannot occur for stationary LSs.

%Figure \ref{fig:two_isolas} shows two figure-eight isolas at $\epsilon=0$, obtained by numerically continuing from two bound states, both obtained in Scenario A for $\epsilon=0.03$, at $r=-0.645$ (narrow isola) and $r=-0.68$ (wide isola) respectively. The wide figure-eight isola corresponds to the type of structure described in \cite{burke2009multipulse}, while the narrow figure-eight isola (which is identical to the thick line in Fig.~\ref{fig:isolas}) has not been described previously in the literature.

%%%%%%%%%%%%%%%%%%%%%%%%%%%%%%%%%%%%%%%%%%%%%%%%%%%%%%%%%%%%%%%%%%%%%%%%%%
\section{Reduced model of interacting localized structures}
\label{sec:reduced_model}
%%%%%%%%%%%%%%%%%%%%%%%%%%%%%%%%%%%%%%%%%%%%%%%%%%%%%%%%%%%%%%%%%%%%%%%%%%%%%%%%%%%%%%%

An important concept in the analysis of LSs is the notion of \textit{spatial dynamics} that applies equally well in the comoving frame. In the simplest case one linearizes the equation of motion (\ref{eq:she_general}) about the trivial state, and considers solutions of the form $u\propto e^{\lambda x}$, cf. \cite{burke2006localized,knobloch2015spatial}. The roots $\lambda$ of the resulting characteristic polynomial, known as \textit{spatial eigenvalues}, govern the behavior of the system in the vicinity of the trivial solution $u\equiv 0$. They are given by
\begin{equation}
    \lambda = \pm\sqrt{\pm\sqrt{r}-1}.
\label{eq:spatial_eigenvalues}
\end{equation}
In the case of interest, $r<0$ so that LSs exist, $\lambda$ is complex: $\lambda = \pm\alpha \pm i \beta$ with $\alpha,\beta>0$. The (positive) real part of $\lambda$ describes the exponential growth of $u$ away from $u=0$ and towards LS on a $1/\alpha$ spatial scale, while the nonzero imaginary part of $\lambda$ implies that this growth is not monotonic, but oscillatory, with wavelength $2\pi/\beta$. For stationary LS the profile decays towards $u=0$ in the same manner.

Interactions between LSs will be mediated by these exponentially growing/decaying oscillatory tails, unless the proximate (i.e.~most strongly interacting) large-amplitude extrema of the two respective structures come close enough for nonlinear effects to become relevant. Such interactions via oscillating tails are found in a wide range of problems and have been studied extensively \cite{coullet1987nature,elphick1991interacting,ei1994equation,balmforth1994chaotic,nishiura2022traveling}. Here, in a spirit similar to these works, in particular \cite{coullet1987nature}, we propose a simple reduced model to quantitatively describe these interactions.

Consider two LSs, with the closest order one amplitude \avk{extrema} of either structure located at $x_i$, $i=1,2$. Each LS is characterized by a drift velocity $c_i$, $i=1,2$, known from the analysis in section \ref{sec:drift_speed}. Note that $c_i$ is zero for symmetric LS, and nonzero for asymmetric LS. The proposed reduced model equations, reminiscent of overdamped particle dynamics, read 
\begin{align}
\frac{dx_1}{dt} &= c_1+ g_1 \cos\left(\beta|x_1-x_2| - \phi\right) e^{-\alpha|x_1-x_2|} \nonumber \\ \frac{dx_2}{dt} &= c_2+g_2 \cos\left(\beta|x_1-x_2| - \phi\right)e^{-\alpha|x_1-x_2|}.
\label{eq:reduced_model_equations}
\end{align}
Equations (\ref{eq:reduced_model_equations}) are integrated using a fourth-order Runge-Kutta method with initial conditions corresponding to DNS data for any given collision. The model involves three unknown parameters describing the interaction: two amplitudes $g_1,g_2$ and a phase $\phi$. The values of these parameters will depend on $r$ and on the colliding structures in question.

Below, we compare the predictions of the reduced model given by Eq.~(\ref{eq:reduced_model_equations}) with high-resolution DNS results using a gradient descent optimization to determine $g_1$, $g_2$ and $\phi$, as described in appendix \ref{sec:app}. To conveniently assess the quantitative agreement between the particle-like trajectories, we consider the deviation of the relative distance between proximate extrema from free propagation. For $x_2>x_1$ (without loss of generality), we define
\begin{equation}
    \chi(t) \equiv x_2(t)-x_1(t) - (c_2 - c_1 )t - [x_2(0)- x_1(0)] \label{eq:def_chi}.
\end{equation}

%%%%%%%%%%%%%%%%%%%%%%%%%%%%%%%%%%%%%%%%%%%
\subsection{Comparison between reduced model and DNS}
\label{ssec:comparison_model_DNS}
%%%%%%%%%%%%%%%%%%%%%%%%%%%%%%%%%%%%%%%%%%%
%%%%%%%%%%%%%%%%%%%%%%%%%%%%%%%%%%%%%%%%%%%
%%%%%%%%%%%%%%%%%%%%%%%%%%%%%%%%%%%%%%%%%%%
\subsubsection{Chasing: scenario A}
\begin{figure}[h]
    \centering
    \includegraphics[width=0.5\textwidth]{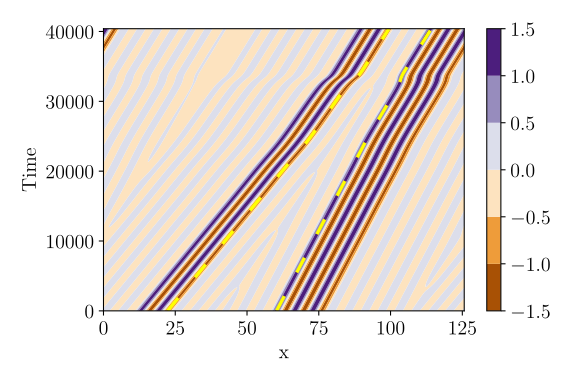}
    \caption{Overlay of the reduced model trajectories on top of DNS data for scenario A. Colored contour plot shows the DNS results at $\epsilon=0.03$, $r=-0.68$ in scenario $A$ (collision between two asymmetric structures that are two and three wavelengths long, respectively). Yellow dashes indicate trajectories predicted by the reduced model (\ref{eq:reduced_model_equations}) with parameters $g_1=0.53489982$, $g_2=-0.30631508$ and $\phi=-1.08804942$, as well as $c_1=0.0020$ and $c_2=0.00128276$. See also Fig.~\ref{fig:chi_scenario_A}. }
    \label{fig:scenario_A_overlay}
\end{figure}

\begin{figure}[h]
\includegraphics[width = 0.5\textwidth]{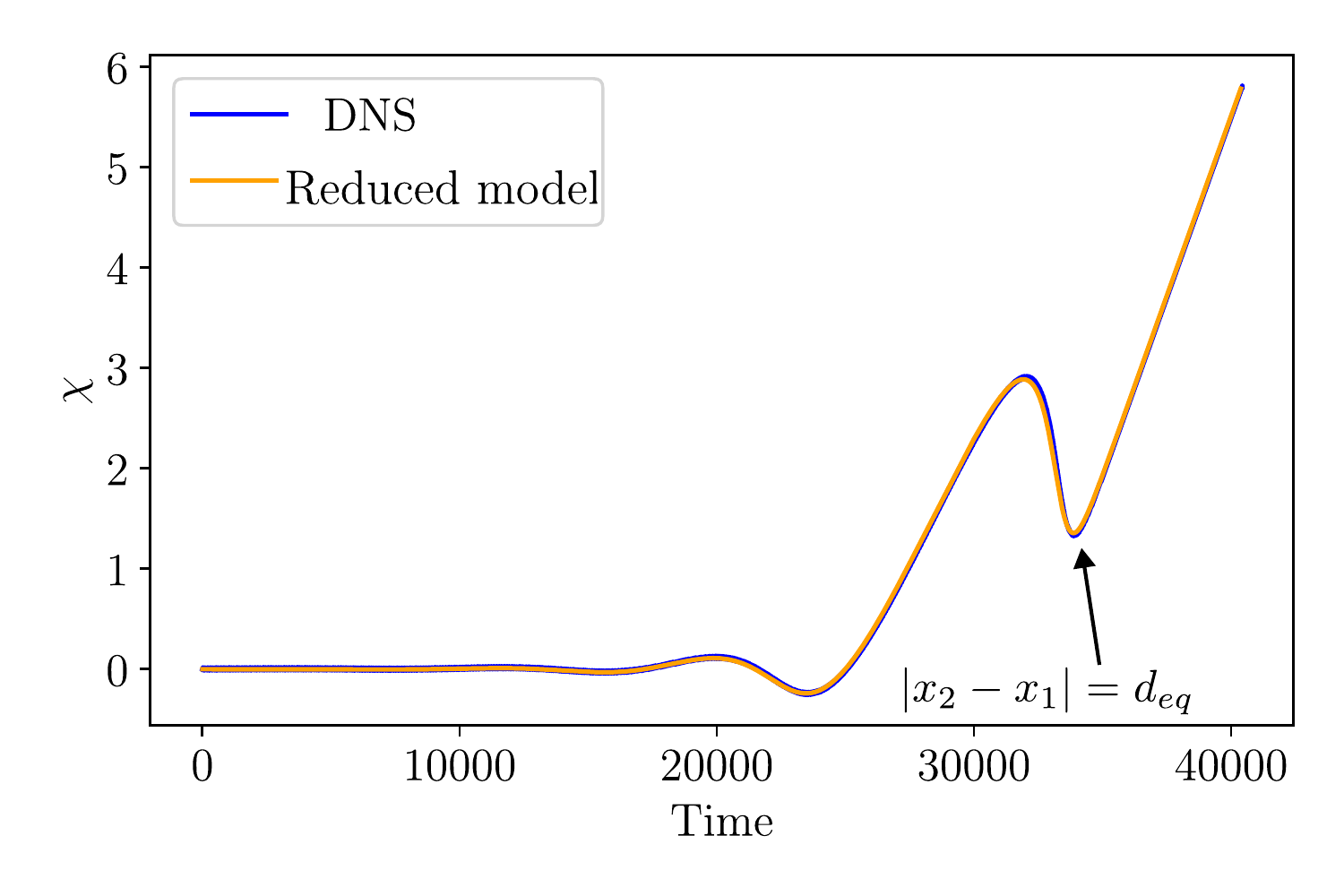}
\caption{Residual distance $\chi$ between proximate extrema (defined in Eq.~(\ref{eq:reduced_model_equations})) versus time from the DNS and the reduced model in scenario A, corresponding to Fig.~\ref{fig:scenario_A_overlay}. A good quantiative agreement is observed between DNS and model. The interaction is effectively repulsive, cf. section \ref{ssec:sign_of_interactions}.}
\label{fig:chi_scenario_A}
\end{figure}

%%%%%%%%%%%%%%%%%%%%%%%%%%%%%%%%%%%%%%%%%%%
Figure \ref{fig:scenario_A_overlay} shows a comparison between the DNS results (contour plot) and the model predictions (dashed yellow lines) for a chasing collision of two asymmetric extrema (scenario A). There is good agreement between the extrema trajectories in both cases, including at a quantitative level, as can be seen from Fig.~\ref{fig:chi_scenario_A}.

%%%%%%%%%%%%%%%%%%%%%%%%%%%%%%%%%%%%%%%%%%%
%%%%%%%%%%%%%%%%%%%%%%%%%%%%%%%%%%%%%%%%%%%%%%%
%%%%%%%%%%%%%%%%%%%%%%%%%%%%%%%%%%%%%%%%%%%%%%%
%%%%%%%%%%%%%%%%%%%%%%%%%%%%%%%%%%%%%%%%%%%%%%%%%%%%
Figure \ref{fig:model_parameters_vs_r} shows the model parameters obtained by the gradient descent method described in appendix \ref{sec:app} at different values of $r$ for the chasing collision shown above. In all cases shown here, a pure bound state is formed in the collision, without the addition or deletion of any extrema. The parameter values are seen to depend only weakly on $r$, reflecting small changes in spatial structure as $r$ is varied for a given type of LS.
%%%%%%%%%%%%%%%%%%%%%%%%%%%%%%%%%%%%%%%%%%%%%%%%%%%%
\begin{figure}
    \centering
    \includegraphics[width=0.5\textwidth]{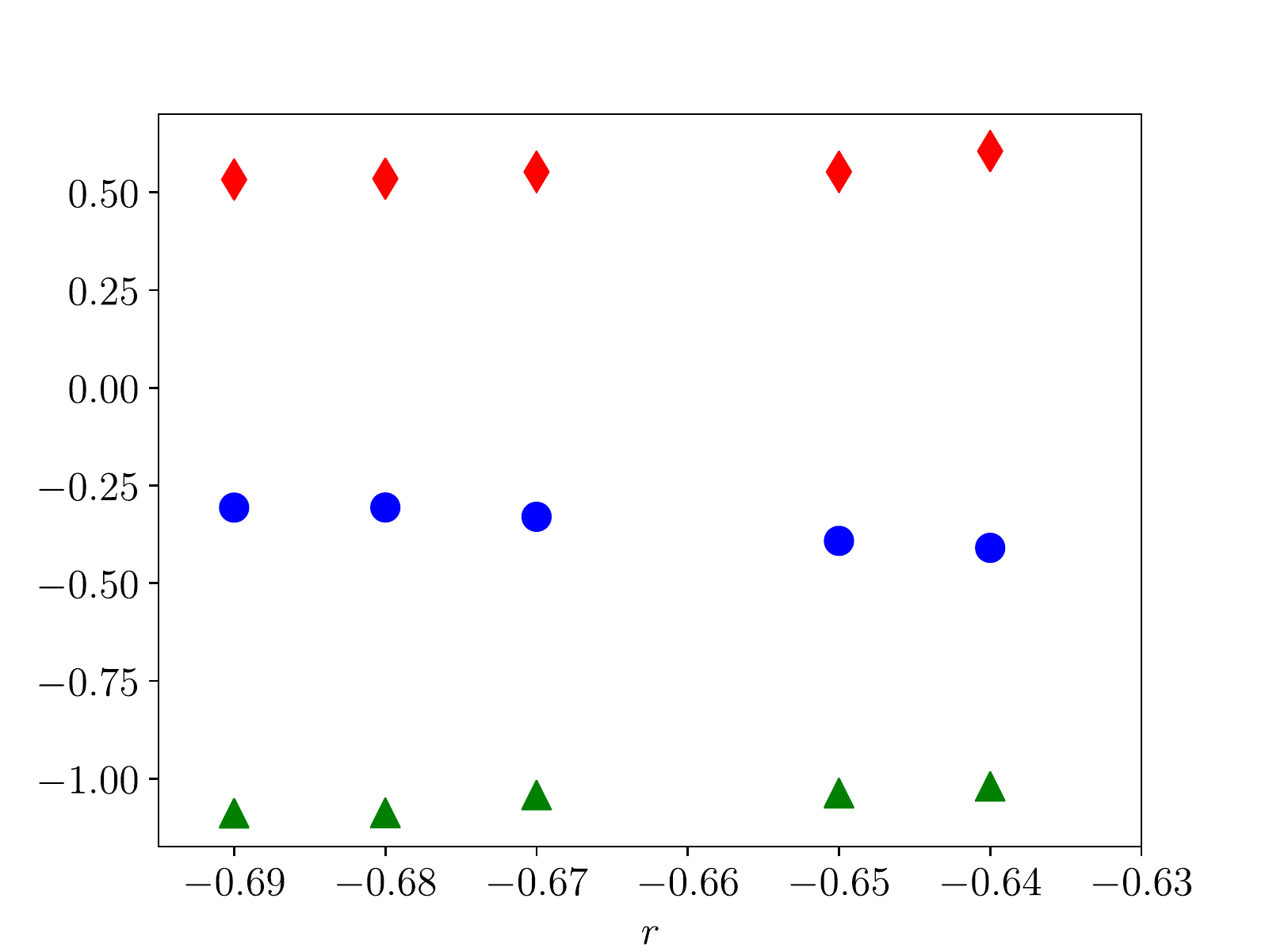}
     \caption{Parameters of the educed model obtained for scenario A and different values of the control parameter $r$. Red diamonds: $g_1$, blue circles: $g_2$, green triangles $\phi$. The parameters depend only weakly on $r$, reflecting minor changes in profile with $r$. All cases shown correspond to collisions generating pure bound states, except $r=-0.65$, where a metastable bound state forms (Fig.~\ref{fig:L2_norm_vs_time}), leading to the appearance of additional extrema at late times. In that special case $r=-0.65$, the fit was performed with a cut-off time $t_{f}$ (cf. appendix \ref{sec:app}) after the initial collision but before the generation of the additional extrema. In the range $-0.67\lesssim r \lesssim -0.65$, and at $r\gtrsim -0.64$, the scenario A collisions do not result in bound states, but instead generate new peaks (cf. Fig.~\ref{fig:Delta_L2_norm}).}
     %and therefore the model does not accurately describe the full collision dynamics, see also Sec.~\ref{sssec:headon_model}.} 
    \label{fig:model_parameters_vs_r}
\end{figure}

%%%%%%%%%%%%%%%%%%%%%%%%%%%%%%%%%%%%%%%%%%%%%%%%%%%%
\subsubsection{Symmetric-asymmetric collision: scenario B}
\begin{figure}[h]
\includegraphics[width = 0.5\textwidth]{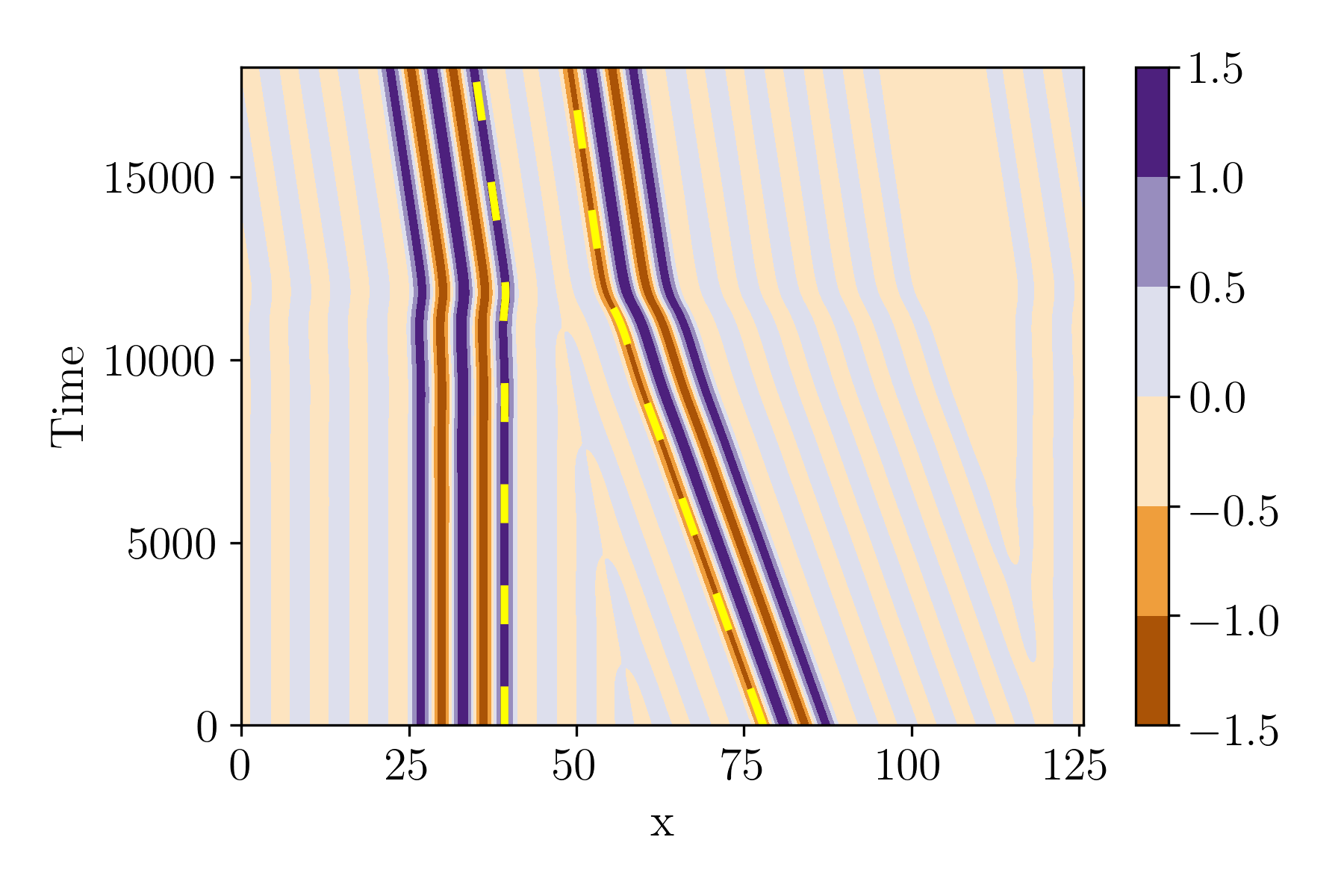}
\caption{Colored contour plot of the DNS of Eq.~(\ref{eq:she_general}) at $\epsilon=0.03$, $r=-0.69$ in scenario $B$ (collision between symmetric and asymmetric LSs, with extrema of opposite sign facing each other). Yellow dashes indicate the trajectories predicted by the reduced model (\ref{eq:reduced_model_equations}) with parameters $g_1=0.3539801$, $g_2=-0.51798143$, and $\phi=-1.038100556$, as well as $c_1=0$ and $c_2=-0.0019615$. See also Fig.~\ref{fig:chi_scenario_B}. }
\label{fig:scenario_B_overlay}
\end{figure}

\begin{figure}[h]
\includegraphics[width = 0.5\textwidth]{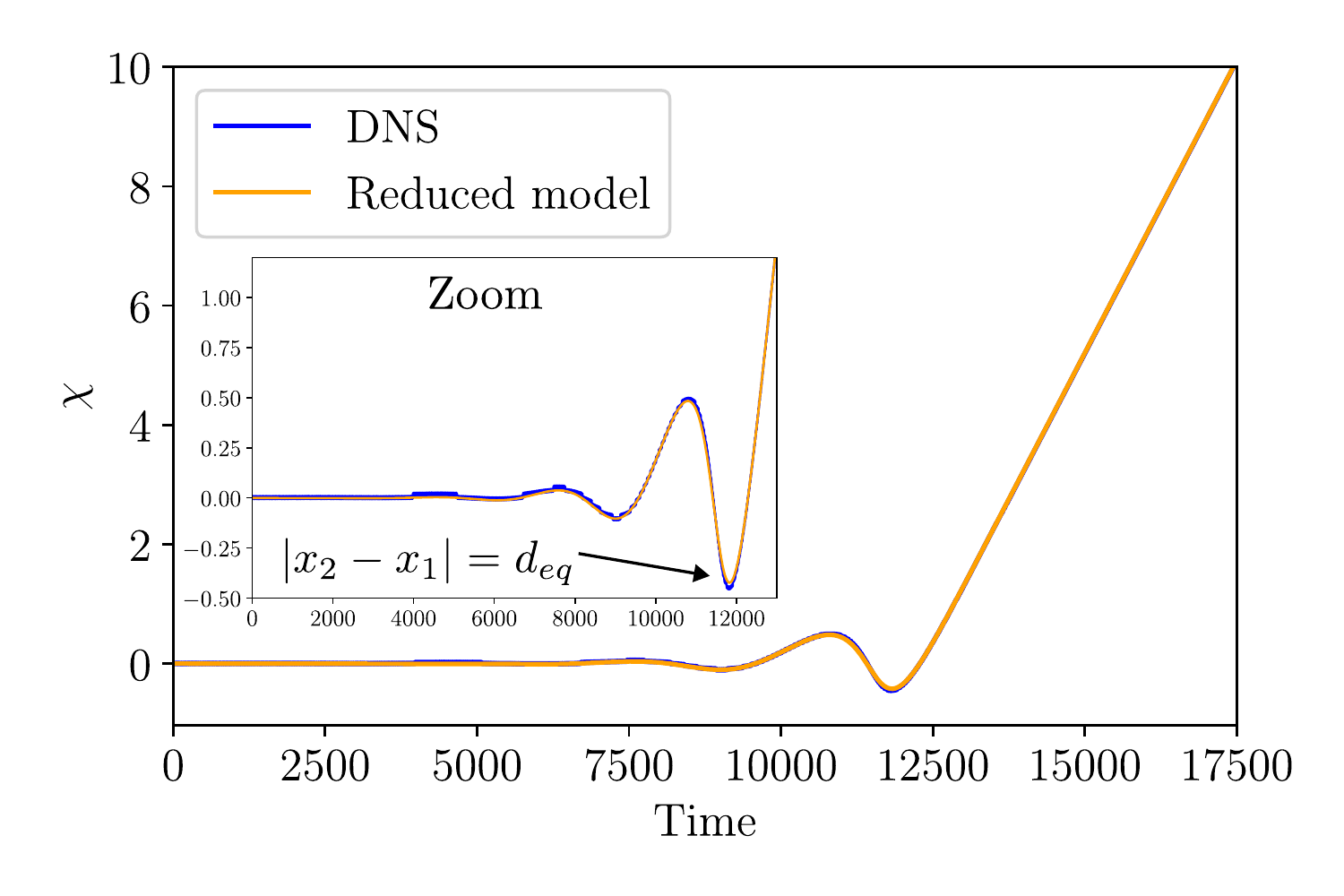}
\caption{Residual distance $\chi$ between proximate extrema (defined in Eq.~(\ref{eq:reduced_model_equations})) versus time from the DNS and the reduced model for scenario B, corresponding to Fig.~\ref{fig:scenario_B_overlay}. The zoom shows the equilibrium separation. The interaction is effectively attractive, cf. section \ref{ssec:sign_of_interactions}. }
\label{fig:chi_scenario_B}
\end{figure}

%%%%%%%%%%%%%%%%%%%%%%%%%%%%%%%%%%%%%%%%%%%%%%%%%%%%%%%
 Figure \ref{fig:scenario_B_overlay} shows an overlay of the reduced model trajectories on top of the DNS results, similar to Fig.~\ref{fig:scenario_A_overlay}, but for a collision between a symmetric LS with a maximum at its center and a two-wavelength asymmetric LS (scenario B). There is again good agreement between \avk{the trajectories of the closest extrema} in both cases, including at a quantitative level, as one can see in terms of the quantity $\chi$ from Fig.~\ref{fig:chi_scenario_B}.
%%%%%%%%%%%%%%%%%%%%%%%%%%%%%%%%%%%%%%%%%%%%%%%%%%%%%%%
%%%%%%%%%%%%%%%%%%%%%%%%%%%%%%%%%%%%%%%%%%%%%%%%%%%%%%%
 %%%%%%%%%%%%%%%%%%%%%%%%%%%%%%%%%%%%%%%%%%%%%%%%%%%%%%%

%%%%%%%%%%%%%%%%%%%%%%%%%%%%%%%%%%%%%%%%%%%%%%%%%%%%%%%
\subsubsection{Flipped symmetric-asymmetric collision: scenario C}
\begin{figure}[h]
    \includegraphics[width=0.5\textwidth]{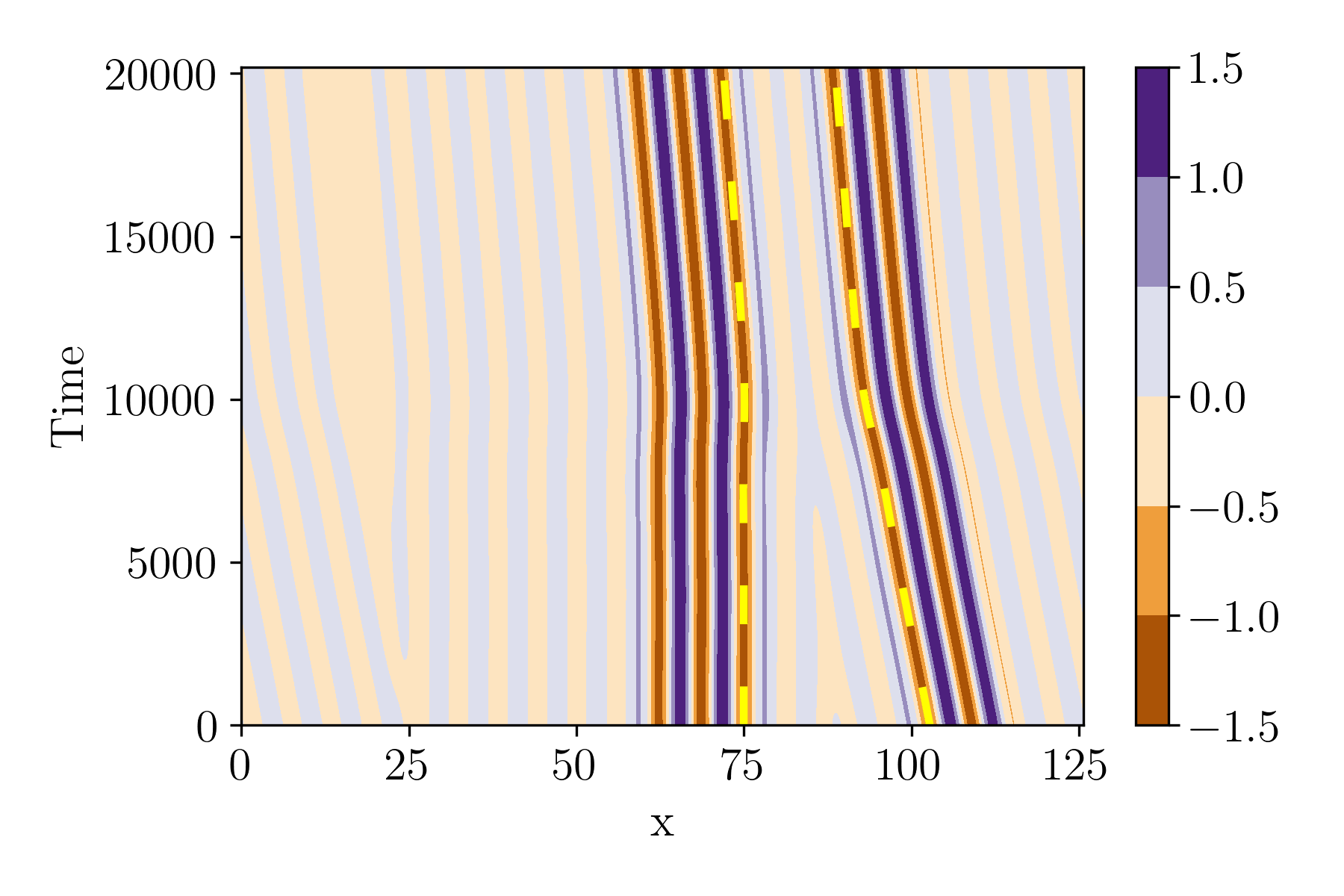}
    \caption{Colored contour plot of the DNS of Eq.~(\ref{eq:she_general}) at $\epsilon=0.03$, $r=-0.645$ in scenario $C$ (collision between symmetric and asymmetric LSs, with extrema of opposite sign facing each other). Yellow dashes indicate the trajectories predicted by the reduced model (\ref{eq:reduced_model_equations}) with parameters $g_1=-0.43756$, $g_2=0.55907$, and $\phi=-1.013089$, as well as $c_1=0$ and $c_2=-0.001895$.
    See also Fig.~\ref{fig:chi_scenario_C}.
    }
    \label{fig:scenario_C_overlay}
\end{figure}

\begin{figure}[h]
    \includegraphics[width=0.5\textwidth]{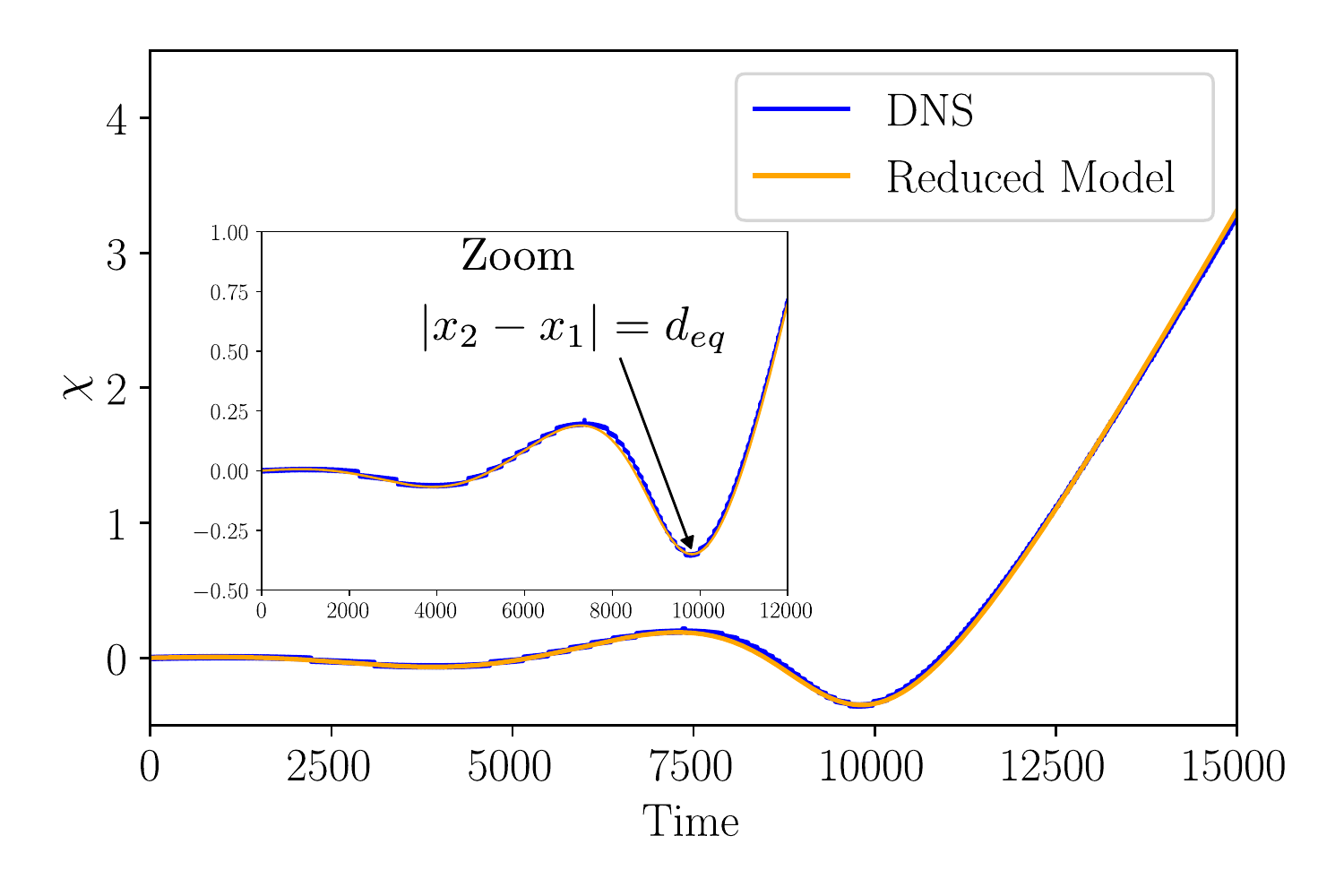}
    \caption{Residual distance $\chi$ between proximate extrema (defined in Eq.~(\ref{eq:reduced_model_equations})) versus time from the DNS and the reduced model for scenario C, corresponding to Fig.~\ref{fig:scenario_C_overlay}. The interaction is effectively attractive, cf. section \ref{ssec:sign_of_interactions}.}
    \label{fig:chi_scenario_C}
\end{figure}

%%%%%%%%%%%%%%%%%%%%%%%%%%%%%%%%%%%%%%%%%%%%%%%%%%%%%%%
 Figure \ref{fig:scenario_C_overlay} shows an overlay of the reduced model trajectories on top of the DNS results, similar to Fig.~\ref{fig:scenario_A_overlay} for a collision from scenario C. There is good agreement between the trajectories of the proximate extrema in both cases, including at a quantitative level, as one can see in terms of the quantity $\chi$ from Fig.~\ref{fig:chi_scenario_C}.

%%%%%%%%%%%%%%%%%%%%%%%%%%%%%%%%%%%%%%%%%%%%%%%%%%%%%%%
\subsubsection{Head-on collision: scenario D}
\begin{figure}[h]
\includegraphics[width = 0.5\textwidth]{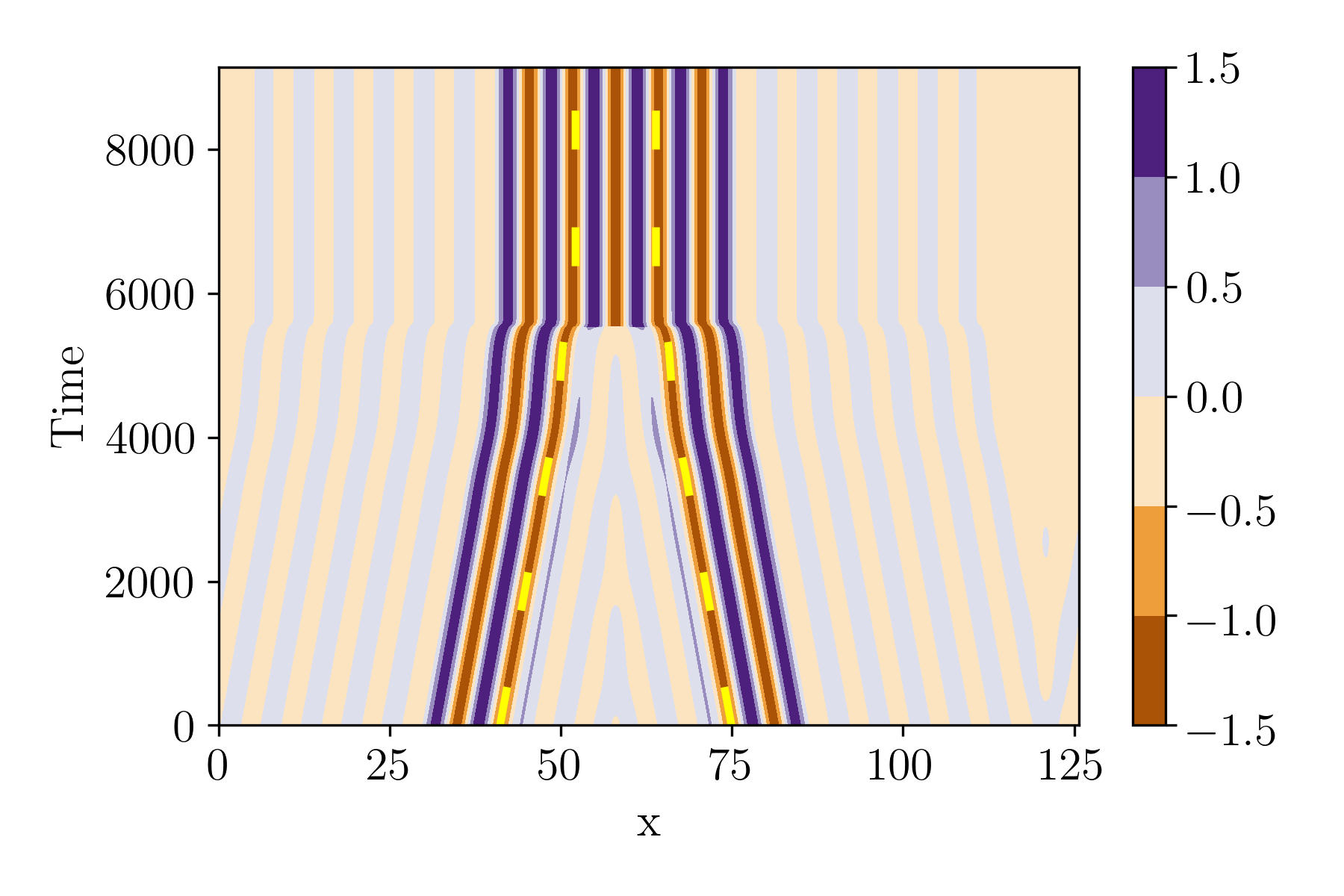}
\caption{Colored contour plot of DNS of Eq.~(\ref{eq:she_general}) at $\epsilon=0.03$, $r=-0.65$ in scenario $D$ (head-on collision of two identical structures). Yellow dashes indicate trajectories predicted by the reduced model (\ref{eq:reduced_model_equations}), with parameters $g_1=-g_2=-0.551$, $\phi=-1.90324$, as well as $c_1=-c_2=0.0019291$. See also Fig.~\ref{fig:chi_scenario_D}. }
\label{fig:scenario_D_overlay}
\end{figure}

\begin{figure}
    \centering
    \includegraphics[width=0.5\textwidth]{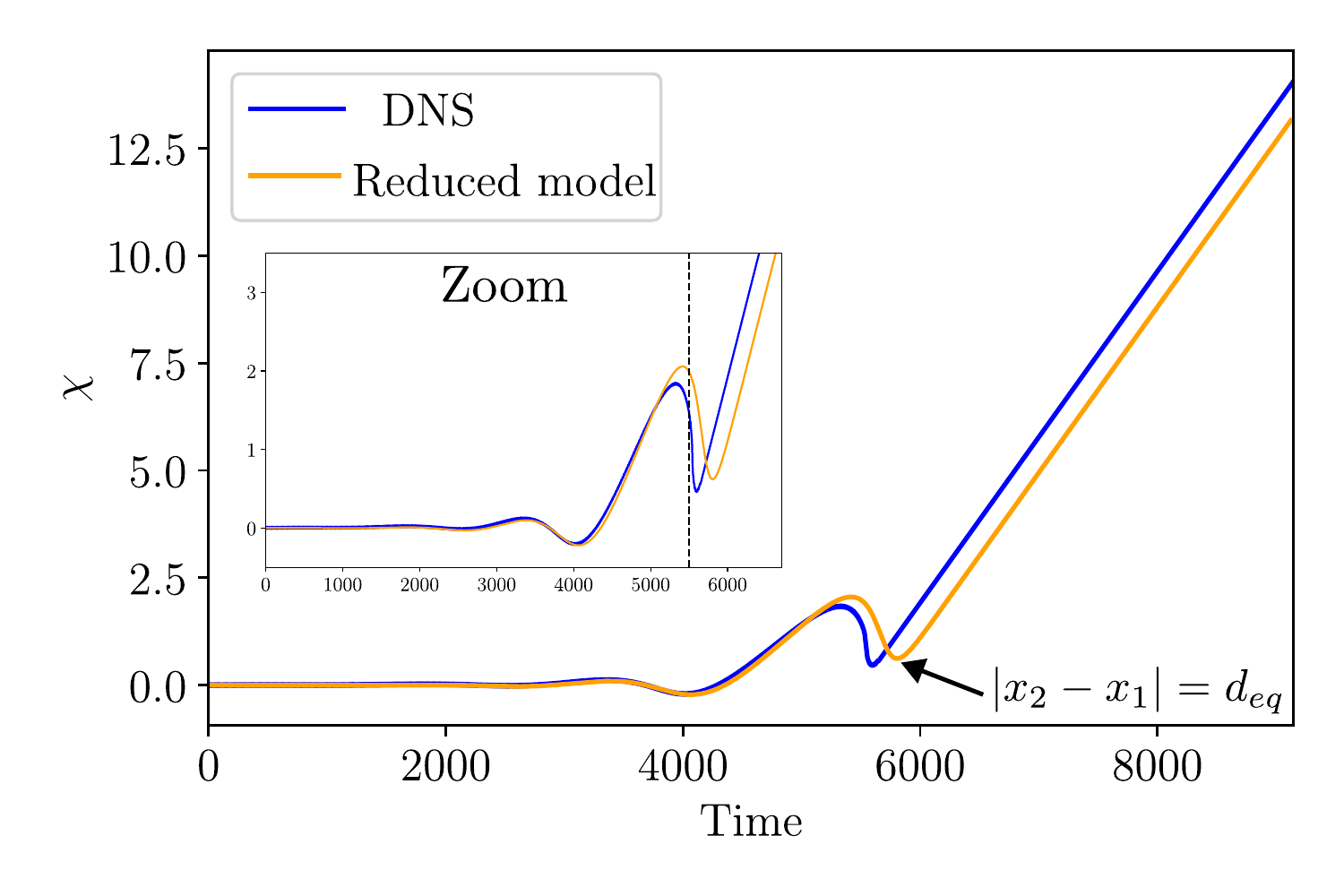}
    \caption{Residual distance $\chi$ between proximate extrema (defined in Eq.~(\ref{eq:reduced_model_equations})) versus time from the DNS and the reduced model, corresponding to Fig.~\ref{fig:scenario_D_overlay}. Inset shows zoom on early times. The vertical dashed line indicates the time $t\approx 5500$ when new extrema appear. The best-fit model prediction deviates from the DNS slightly before this, as a consequence of nonlinear effects not captured by the model.}
    \label{fig:chi_scenario_D}
\end{figure}

%%%%%%%%%%%%%%%%%%%%%%%%%%%%%%%%%%%%%%%%%%%%%%%%%%%%%%%
\label{sssec:headon_model}
 Figure \ref{fig:scenario_D_overlay} shows an overlay of the reduced model trajectories on top of the DNS results for a collision from scenario D. While the trajectories in Fig.~\ref{fig:scenario_D_overlay} look qualitatively consistent, the quantitative perspective provided by Fig.~\ref{fig:chi_scenario_D} reveals that the reduced model with optimal parameters clearly deviates from the DNS results in this case. The gradient descent method was applied with $t_{f}=5~500$ (see appendix~\ref{sec:app}) to find the set of parameters used in Fig.~\ref{fig:chi_scenario_D}, but no better agreement was found for other choices of $t_{f}$, which were also tested. 

The deviation between the reduced model trajectories and the DNS result is in agreement with expectations. Since the reduced model is exclusively built on the linear structure of SH35, while the creation of new extrema is a highly nonlinear process, the reduced model is not expected to capture the dynamics correctly once nonlinear interactions become dominant. At earlier times, the values of $\chi$ produced by the reduced model remain close to the DNS data, and only deviate near the time of extremum creation. This highlights the limitations of the otherwise very successful reduced model analysed here.

%We use the set of coupled equations with $r=-0.65$, $c = 0.001925$ (from DNS which slightly deviates from the asymptotic theory as shown earlier in the paper), and a cutoff at $x = 5000$ for our gradient descent algorithm  and allow it to converge to an optimal state. 

%From our descent algorithm, we get $g_1 = g_2 = 0.534$ and a phase shift of $\phi = 4.853$ and the following result with the trajectories of the particle model mapped on top of the DNS results.

 %%%%%%%%%%%%%%%%%%%%%%%%%%%%%%%%%%%%%%%%%%%%%%%%%%%%%%%
%%%%%%%%%%%%%%%%%%%%%%%%%%%%%%%%%%%%%%%%%%%%%%%%%%%%%%%
%%%%%%%%%%%%%%%%%%%%%%%%%%%%%%%%%%%%%%%%%%%%%%%%%%%%%%%

%%%%%%%%%%%%%%%%%%%%%%%%%%%%%%%%%%%%%%%%%%%%%%%%%%%%%%%
\subsection{Sign of the interaction: attractive or repulsive?} 
\label{ssec:sign_of_interactions}
\begin{figure}
    \centering
    \includegraphics[width=0.5\textwidth]{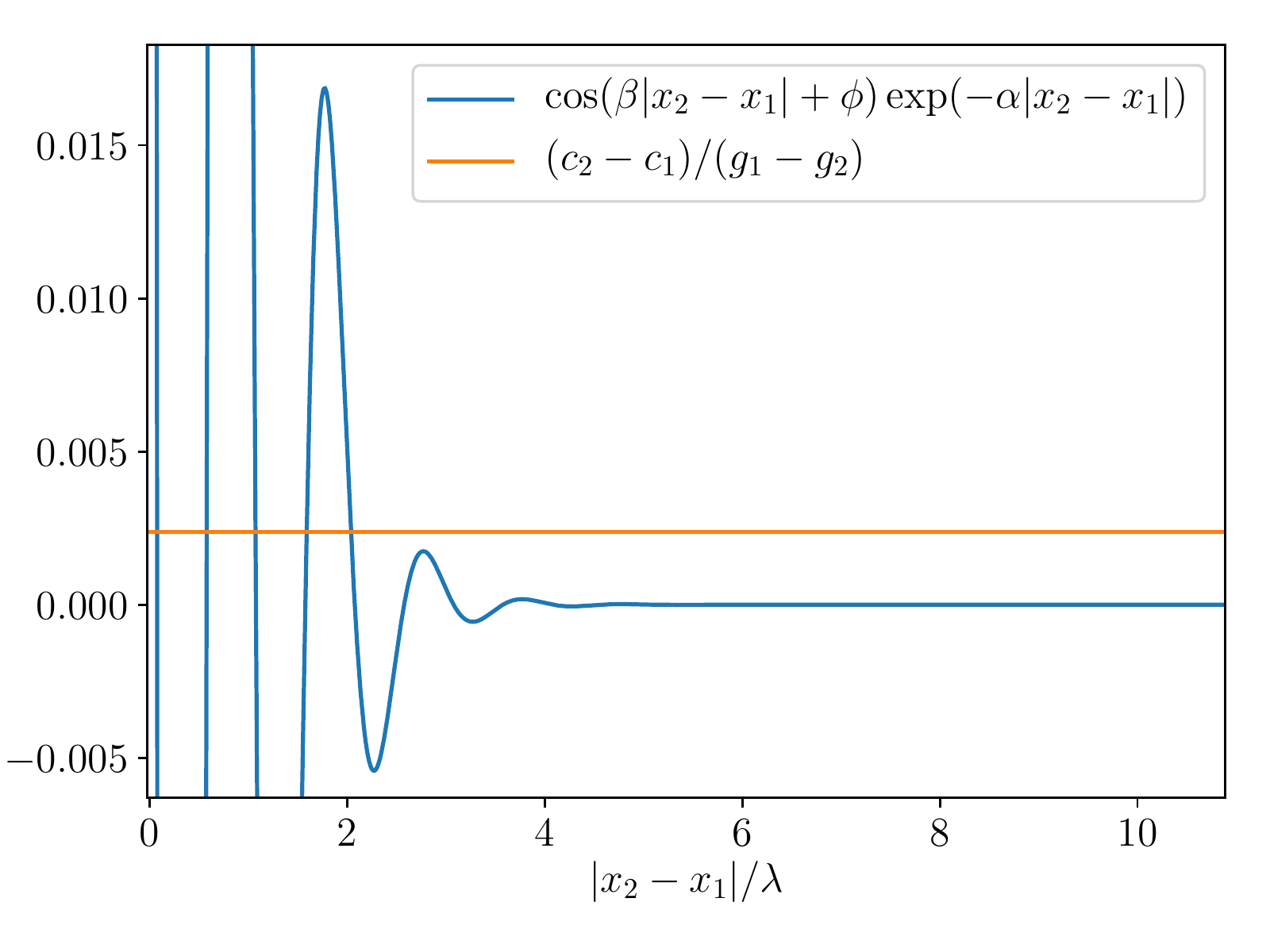}
    \caption{Illustration of how the equilibrium distance $d_{eq}$ is determined from Eq.~(\ref{eq:def_deq}).}
    \label{fig:def_deq}
\end{figure}

%%%%%%%%%%%%%%%%%%%%%%%%%%%%%%%%%%%%%%%%%%%%%%%%%%%%%%%
It is interesting to note that in all the cases described above, when the interacting extrema were of opposite signs (scenarios A, B), we found $g_1>0$ and $g_2<0$ (for $-\pi <\phi < 0$). Conversely, when the two interacting extrema were of the same sign (scenarios C, D), we systematically found $g_1<0$ and $g_2>0$ (for $-\pi <\phi < 0$). However, it is nontrivial to deduce what this implies for the sign of the effective interaction, due to its oscillatory nature. We can define an effective sign of the interaction as follows. First we introduce the \textit{equilibrium distance} $d_{eq}$ as the value of $|x_2-x_1|$ in the bound state after the collision has occurred. From Eq.~(\ref{eq:reduced_model_equations}), we see that equilibrium implies
\begin{equation}
    \cos\left(\beta d + \phi \right)\exp(-\alpha d) = \frac{c_2-c_1}{g_1-g_2}.
    \label{eq:def_deq}
\end{equation}
In general, Eq.~(\ref{eq:def_deq}) has multiple solutions, as illustrated in Fig.~\ref{fig:def_deq}. One may surmise that these multiple solutions are related to the overlapping isolas shown in Fig.~\ref{fig:two_isolas}, which differ in their equilibrium distance by one wavelength. When $c_2=c_1$, Eq.~(\ref{eq:def_deq}) has infinitely many solutions, similar to the family of bound two-pulse states described in \cite{burke2009multipulse} that differ only in their equilibrium separation. By contrast, for $c_2-c_1 \neq 0$, the number of solutions is finite. In a generic collision at $\epsilon>0$, since the colliding LSs approach from a large distance, we expect that the physical value of $d_{eq}$ is given by the largest positive value of $d$ that solves Eq.~(\ref{eq:def_deq}). 
%%%%%%%%%%%%%%%%%%%%%%%%%%%%%%%%%%%%%%%
%%%%%%%%%%%%%%%%%%%%%%%%%%%%%%%%%%%%%%%

We assume that $d_{eq}$ is known for a given collision of two LSs starting out at a distance $|x_2-x_1|=d_0$, with $c_1>c_2$ (which applies to all examples shown above). In the absence of interactions, the free propagation of the two LSs will reduce the distance from $d_0$ to $d_{eq}$ in a time
\begin{equation}
    t_{free} = \frac{d_0-d_{eq}}{c_1-c_2}.
\end{equation}
In the presence of interactions, this time will change to $t_{eq}$, a time that may be larger or smaller than $t_{free}$ due to the oscillatory nature of the interaction. 

We propose the following terminology. If the equilibrium distance in the collision is reached \textit{earlier} than for free propagation, i.e. if $t_{eq} < t_{free}$, then we say that the interaction is \textit{effectively attractive}. By contrast, if the equilibrium distance is reached \textit{later}, i.e. $t_{eq} > t_{free}$, then we say that the interaction is \textit{effectively repulsive}.  \\
We note that the effective sign of the interaction can be determined graphically from the sign of $\chi(t_{eq})$, since Eq.~(\ref{eq:def_chi}) implies
\begin{equation}
    t_{eq} = t_{free} + \frac{\chi(t_{eq})}{|c_1-c_2|}.
\end{equation}
The time $t=t_{eq}$, where the distance $|x_2-x_1|=d_{eq}$ is highlighted by an arrow in Figs.~\ref{fig:chi_scenario_A}, \ref{fig:chi_scenario_B}, \ref{fig:chi_scenario_C} and \ref{fig:chi_scenario_D}; since $x_2-x_1$ is constant at $t>t_{eq}$, a linear increase in $\chi=const+ (c_1-c_2) t$ sets in.  %The time $t_{eq}$ is highlighted by an arrow in Figs.~\ref{fig:chi_scenario_A}, \ref{fig:chi_scenario_B} and \ref{fig:chi_scenario_D} by an arrow.
The sign of $\chi(t_{eq})$, i.e. the onset of this linear increase, determines the effective sign: if $\chi(t_{eq})>0$, then the interaction is effectively repulsive but if $\chi(t_{eq})<0$, then the interaction is effectively attractive.

We highlight that even though the collisions from scenarios A and B shown in Figs.~\ref{fig:scenario_A_overlay} and \ref{fig:scenario_B_overlay} both feature interactions between extrema of opposite signs, the relative signs of their interaction differ: the interaction is effectively repulsive in scenario A (Fig.~\ref{fig:chi_scenario_A}), while it is effectively attractive in scenario B (Fig.~\ref{fig:chi_scenario_B}). This indicates that while the relative sign of interacting extrema does appear to be important, it is not the only factor determining collision outcomes. This is likely related to the observed deviation between scenarios A, B and C, D, respectively, in the parameter range $-0.65 \lesssim r \lesssim -0.64$.

%%%%%%%%%%%%%%%%%%%%%%%%%%%%%%%%%%%%%%%%%%%%%%%%%%%%%%%
%%%%%%%%%%%%%%%%%%%%%%%%%%%%%%%%%%%%%%%%%%%%%%%%%%%%%%%
\section{Conclusions}
\label{sec:conclusions}
%%%%%%%%%%%%%%%%%%%%%%%%%%%%%%%%%%%%%%%%%%%%%%%%%%%%%%%%%
In this paper, we have provided an in-depth analysis of LSs in the nonvariational SH35, Eq.~(\ref{eq:she_general}), using numerical continuation, DNSs, asymptotics and reduced-order modeling to shed new light on the propagation and interaction of LSs in a canonical driven dissipative system. These interactions are highly inelastic, in contrast to systems described by integrable partial differential equations, and lead to both stationary and drifting structures. Moreover, the interactions can be attracting or repelling depending on the nature of the interacting LSs and the parameters.

The asymptotic theory predicts a linear dependence of the drift speed of LSs on $\epsilon$, with excellent quantitative agreement with DNSs and numerical continuation at $\epsilon \lesssim 0.1$, but significant deviation for $\epsilon\gtrsim 0.1$. The collisions resulting from this drift are shown to display rich phenomenology: different numbers of extrema can be added or deleted in a collision, depending on the types of structures interacting and the value of $r$. Alternatively, a pure bound state can be formed, which preserves the number of extrema of the initial structures. We have found that the stability properties of these bound states play a key role in determining whether a collision changes the number of extrema or not: if the bound state exists stably, and has a sufficiently large basin of attraction, then the perturbation resulting from a collision does not change the number of extrema. However, if this is not the case, then \avk{extrema} will be added or deleted in the collision. %A notable pattern is that, except near the edges of the range of existence of stable propagating solutions, whenever peaks are added the added number of peaks is odd if the two closest peaks are of the same sign, and odd if they are of opposite signs.
When extrema are deleted, this can lead to smaller LSs, or to annihilation in the case of the minimal, single-wavelength asymmetric structure. When extrema are created, metastable states may arise from collisions, which undergo a further nonlinear interaction at late times. This phenomenology is reminiscent of what has been attributed to a \textit{scattor} \cite{nishiura2003dynamic,nishiura2005scattering}: unstable stationary or time-periodic patterns which direct orbits during the collision process in the infinite-dimensional state space along their stable and unstable manifolds. While these ideas were proposed in the context of models other than SH35, they may be applicable to some of the collision events observed here.

%There is a general trend for the number of peaks created in any collision scenario to increase, with increasing $r$, with the only deviations occurring in a small interval $-0.65 \lesssim r \lesssim -0.64$ \avk{(Can we explain this in terms of the range of stability on the bound state isolas? -> ADD DISCUSSION)}. The number of peaks created at a given value of $r$ is strongly correlated with the relative sign of interacting peaks: in scenarios A,B two peaks of opposite signs interact, while in scenarios C,D, same-sign peaks interact. 

Bound states arising from collisions were shown to lie on isolas which are embedded within the snakes and ladders bifurcation structure. In contrast with the variational case $\epsilon=0$, the isolas at $\epsilon>0$ are typically not of figure-eight form, but rather form a complex, spaghetti-like tangled set. Only for the specific subclass of collisions between chasing asymmetric LSs, are the resulting multi-pulse bound states found to lie on isolas whose figure-eight form is preserved at $\epsilon>0$.
%In addition, we believe to have found a new type of narrow figure-eight isola structure existing at $\epsilon=0$ which does not span the entire snaking region, which has not been described in the literature previously.
In addition, we have also described a novel example of an isola at $\epsilon>0$ which is not closed, but features T-point bifurcations at either end, each of which involves the gradual separation of a bound state into two separate asymmetric LSs, each separately corresponding to a homoclinic connection to $u=0$ in the comoving frame. States of this type require drift and so can only be found in nongradient systems of the type studied here.

Finally, a reduced model was also proposed, consisting of two coupled ordinary differential equations, describing the interaction of LSs via their oscillatory exponential tails. The reduced model was shown to reproduce a wide range of collisions, with quantitative accuracy, provided no large-amplitude extrema are created or destroyed in the collision. If the number of large-amplitude extrema is altered in the collision, the model still describes the trajectories of the interacting proximate extrema up until shortly before the time at which this occurs. After this time, the model fails to describe the DNS results since it does not include the nonlinear structure of Eq.~(\ref{eq:she_general}). The fact that the trajectories of LSs can be accurately reproduced by the reduced model for a sizable fraction of all observed cases is indicative of the fact that the collisions between localized patterns are to a large extent particle-like, with a nontrivial interaction potential corresponding to the linear structure of Eq.~(\ref{eq:she_general}). This model led to considerable insight into the nature of the interactions between LSs in the SH35 model and allowed a relatively simple determination of the conditions under which the interaction is effectively attracting or repelling.

Collisions of convectons in binary fluid convection, such as those described in \cite{mercader2013travelling}, feature dynamics beyond the scope of the order parameter description provided by SH35, in particular due to the possibility of complex temporal behavior. Nonetheless, it remains to be understood to what extent the SH35 collision phenomenology described here can be observed between convectons or similar structures in other continuum systems. Specifically, \cite{mercader2013travelling} do not observe bound states, while these play an important role in the selection of collision outcomes described here. It remains an open question whether such bound states exist between convectons, given that binary fluid convectons interact nonlocally via a large-scale solute field generated by the pumping mechanism described in \cite{batiste2006spatially}.

A simple possibility for obtaining more complex temporal behavior in SH35 would be the inclusion of higher-order time-derivatives, e.g. a second-order term that allows for inertia and wave-like solutions. In this case, the interactions described here, which rely exclusively on the exponential tails of LSs, would be enriched due to the possibility of wave radiation. The study of this modified SH35 with higher-order time derivatives is left for a future study.

%%%%%%%%%%%%%%%%%%%%%%%%%%%%%%%%%%%%%
\begin{acknowledgments}
%%%%%%%%%%%%%%%%%%%%%%%%%%%%%%%%%%%%%
We acknowledge support from the National Science Foundation under grants DMS-1908891 and DMS-2009563. We also thank Nicolás Verschueren van Rees for fruitful discussions and for providing us with a high performance C++ solver which we adapted for our DNS studies.
\end{acknowledgments}

\appendix 
%%%%%%%%%%%%%%%%%%%%%%%%%%%%%%%%%%%%%%%%%%%%%%%%%%%%%%%%%%
\section{Reduced model parameter estimation by gradient descent}
\label{sec:app}
%%%%%%%%%%%%%%%%%%%%%%%%%%%%%%%%%%%%%%%%%%%%%%%%%%%%%%%%%%
\begin{figure}
    \centering
    \includegraphics[width=0.5\textwidth]{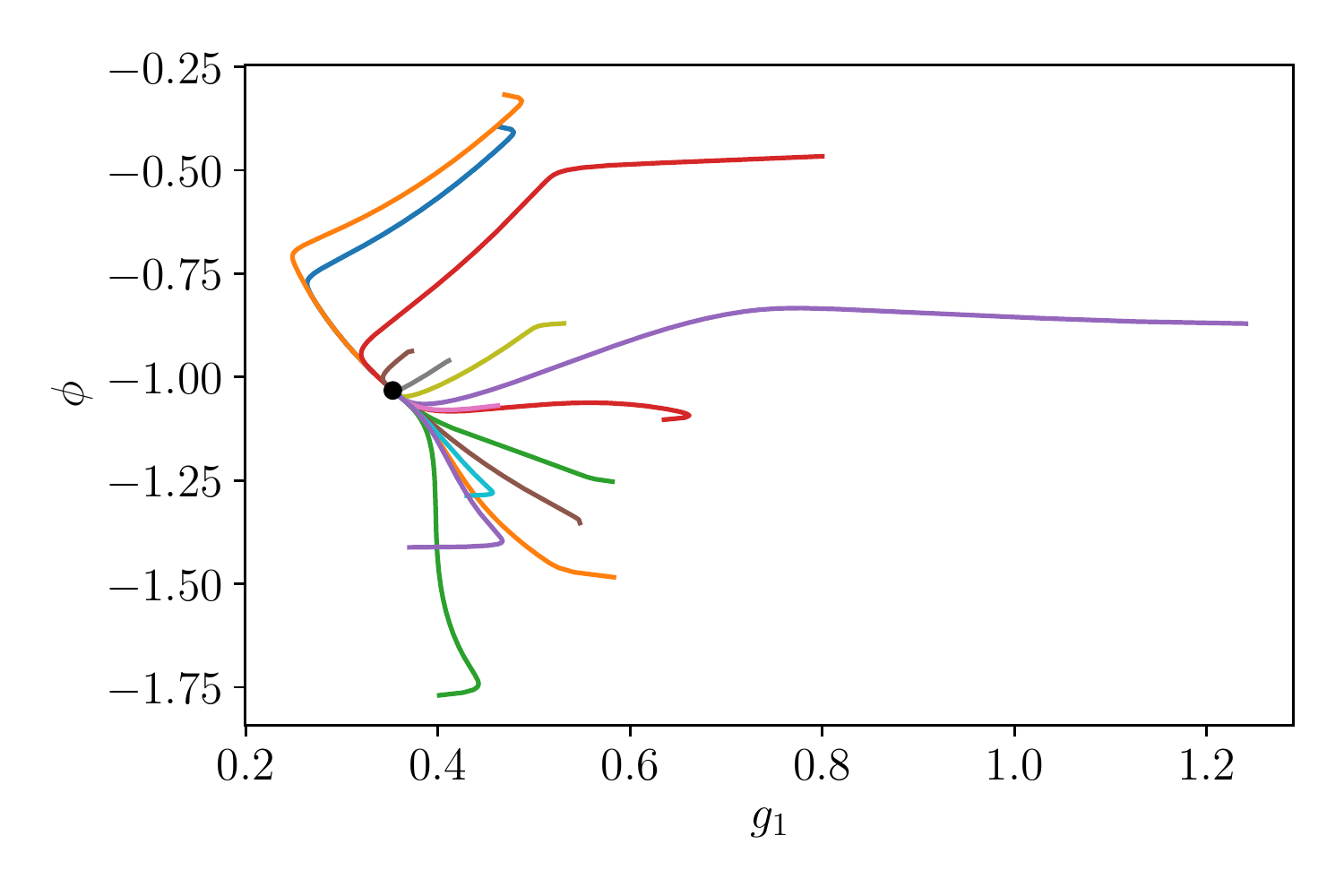}
        \includegraphics[width=0.5\textwidth]{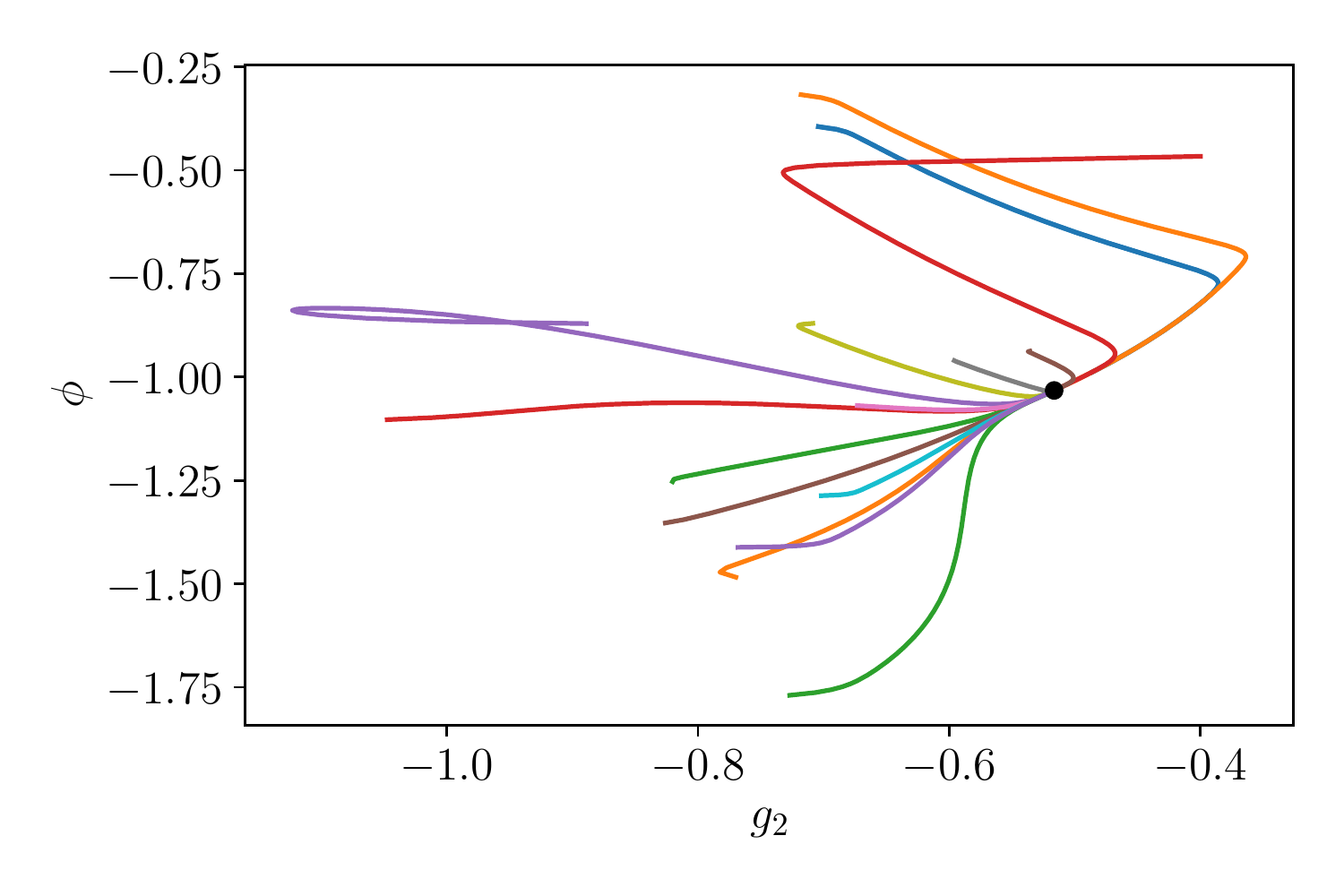}
    \caption{Illustration of the robustness of the gradient descent method: starting from various different initial guesses for the model parameters the method converges to a unique set of parameters $g_1$, $g_2$, $\phi$. Example shown corresponds to scenario $B$ at $r=-0.69$.}
    \label{fig:gradient_descent_robust}
\end{figure}
In this appendix, we describe the details of the gradient descent method used to systematically determine the reduced model parameters $g_1$, $g_2$ and $\phi$ corresponding to SH35 solutions for a given $r$ and a given collision scenario, based on high-resolution DNS data.

Given $u(x,t)$ from DNS on a uniform grid with $8192$ points, we track the position of the two extrema that are located closest to one another in the collision, as a function of time $t$ (note that a high spatial resolution is required for well-resolved tracking). We denote the positions of the closest extrema by $x_1^{DNS}(t)$, $x_2^{DNS}(t)$, corresponding to $x_1(t)$, $x_2(t)$ in Eq.~(\ref{eq:reduced_model_equations}). To determine which set of reduced model parameters $g_1$, $g_2$, $\phi$ most accurately represents the DNS results, we define a cost function
\begin{equation}
    C_{g_1,g_2,\phi}[x_1,x_2] \equiv \sqrt{\frac{1}{t_{f}} \int_0^{t_{f}} \sum_{i=1}^2 \left(x_i(t)-x_i^{DNS}(t)\right)^2 dt},
\end{equation}
where at $t=0$ the colliding LSs are separated by many wavelengths, and $t_{f}$ can be chosen to be long after the collision, if no large-amplitude extrema are created or destroyed (e.g. Fig.~\ref{fig:scenario_A_overlay}), or instead may be chosen just before nonlinear effects set in (e.g. Fig.~\ref{fig:chi_scenario_D}). The global minimum of the functional $C_{g_1,g_2,\phi}[x_1,x_2]$ corresponds to the best fit between reduced model and DNS.

To perform the gradient descent optimization, we start with an initial guess for $g_1$, $g_2$, $\phi$, and compute the gradient $\nabla C = (\partial C/\partial g_1, \partial C/\partial g_2, \partial C /\partial \phi)$. Then, in each step, we update parameters by the rule
\begin{equation}
    (g_1,g_2,\phi) \to   (g_1,g_2,\phi)-\lambda\nabla C,
\end{equation}
with a parameter $\lambda$ measuring the step size in parameter space. 

Figure \ref{fig:gradient_descent_robust} shows sample trajectories in parameter space (top panel: projection onto the $g_1,\phi$ plane, bottom panel: projection onto the $g_2,\phi$ plane) obtained by this method for a collision from scenario $B$ (asymmetric LS colliding with a stationary symmetric LS) at $r=-0.69$ (collision results in pure bound state). For a wide range of initial conditions, the method converges to a well-defined set of parameters. The resulting model trajectories accurately reproduce those observed in the DNS, as can be seen in Figs.~\ref{fig:scenario_A_overlay} and \ref{fig:chi_scenario_A}. The significance of Fig.~\ref{fig:gradient_descent_robust} is that it indicates that the method is robust with respect to the precise initial guess for the model parameters $g_1,g_2,\phi$. However, several complicating factors need to be taken into account to obtain the correct optimal model trajectories with this procedure. 
\begin{enumerate}
\item Due to the very long simulation times of $O(10^4-10^5)$ time units, the propagation velocities $c_1,c_2$ have to be specified with high precision in order to accurately describe the free propagation of the LSs. Only after carefully determining $c_1,c_2$ with a high precision does the gradient descent algorithm converge to an accurate fit between model and DNS.
\item Equations (\ref{eq:reduced_model_equations}) are invariant under $\phi\to \phi+2\pi$ and $\phi\to\phi+\pi$, $(g_1,g_2)\to (-g_1,-g_2)$. The global minimum of $C$ is only unique if the phase is limited to an interval of size $\pi$. Here, we choose $\phi\in[-\pi,0]$.
\item The cut-off time $t_{f}$ can affect the quality of the fit: if it is too large, the final bound state will be weighted excessively compared to the actual collision dynamics of interest. If $t_{f}$ is too short, it may fail to correctly capture the later stages of the collision. We typically choose $t_{f}$ to be several thousand time units after the equilibrium distance is reached, for pure bound states, and before the nonlinear creation of new extrema, if any.
\end{enumerate}

\bibliographystyle{unsrt}
\bibliography{main}

\begin{thebibliography}{10}

\bibitem{korteweg1895xli}
Diederik~Johannes Korteweg and Gustav De~Vries.
\newblock {X}{L}{I}. {O}n the change of form of long waves advancing in a
  rectangular canal, and on a new type of long stationary waves.
\newblock {\em The London, Edinburgh, and Dublin Philosophical Magazine and
  Journal of Science}, 39(240):422--443, 1895.

\bibitem{laing2002multiple}
Carlo~R Laing, William~C Troy, Boris Gutkin, and G~Bard Ermentrout.
\newblock Multiple bumps in a neuronal model of working memory.
\newblock {\em SIAM J. Appl. Math.}, 63(1):62--97, 2002.

\bibitem{ghorayeb1997double}
Kassem Ghorayeb and Abdelkader Mojtabi.
\newblock Double diffusive convection in a vertical rectangular cavity.
\newblock {\em Phys. Fluids}, 9(8):2339--2348, 1997.

\bibitem{batiste2006spatially}
Oriol Batiste, Edgar Knobloch, Arantxa Alonso, and Isabel Mercader.
\newblock Spatially localized binary-fluid convection.
\newblock {\em J. Fluid Mech.}, 560:149--158, 2006.

\bibitem{gad1981growth}
Mohamed Gad-El-Hak, Ron~F Blackwelderf, and James~J Riley.
\newblock On the growth of turbulent regions in laminar boundary layers.
\newblock {\em J. Fluid {M}ech.}, 110:73--95, 1981.

\bibitem{schneider2010localized}
Tobias~M Schneider, Daniel Marinc, and Bruno Eckhardt.
\newblock Localized edge states nucleate turbulence in extended plane {C}ouette
  cells.
\newblock {\em J. Fluid {M}ech.}, 646:441--451, 2010.

\bibitem{coullet2000stable}
Pierre Coullet, C~Riera, and Charles Tresser.
\newblock Stable static localized structures in one dimension.
\newblock {\em Phys. {R}ev. {L}ett.}, 84(14):3069, 2000.

\bibitem{yochelis2008formation}
Arik Yochelis, Yin Tintut, Linda~L Demer, and Alan Garfinkel.
\newblock The formation of labyrinths, spots and stripe patterns in a
  biochemical approach to cardiovascular calcification.
\newblock {\em New J. Phys.}, 10(5):055002, 2008.

\bibitem{knobloch2015spatial}
Edgar Knobloch.
\newblock Spatial localization in dissipative systems.
\newblock {\em Annu. Rev. Condens. Matter Phys.}, 6(1):325--359, 2015.

\bibitem{hohenberg1992effects}
Pierre~C Hohenberg and Jack~B Swift.
\newblock Effects of additive noise at the onset of {R}ayleigh-{B}{\'e}nard
  convection.
\newblock {\em Phys. Rev. A}, 46(8):4773, 1992.

\bibitem{ma2011diagrammatic}
Yi-Ping Ma and Edward~A Spiegel.
\newblock A diagrammatic derivation of (convective) pattern equations.
\newblock {\em Physica D: Nonlinear Phenomena}, 240(2):150--165, 2011.

\bibitem{burke2006localized}
John Burke and Edgar Knobloch.
\newblock Localized states in the generalized {S}wift-{Hohenberg} equation.
\newblock {\em Phys. Rev. E}, 73(5):056211, 2006.

\bibitem{burke2007snakes}
John Burke and Edgar Knobloch.
\newblock Snakes and ladders: localized states in the {S}wift--{H}ohenberg
  equation.
\newblock {\em Phys. Lett. A}, 360(6):681--688, 2007.

\bibitem{burke2007homoclinic}
John Burke and Edgar Knobloch.
\newblock Homoclinic snaking: structure and stability.
\newblock {\em Chaos}, 17(3):037102, 2007.

\bibitem{kozyreff2007nonvariational}
Gregory Kozyreff and Mustapha Tlidi.
\newblock Nonvariational real {S}wift-{H}ohenberg equation for biological,
  chemical, and optical systems.
\newblock {\em Chaos}, 17(3):037103, 2007.

\bibitem{burke2012localized}
John Burke and Jonathan~HP Dawes.
\newblock Localized states in an extended {S}wift--{H}ohenberg equation.
\newblock {\em SIAM J. Appl. Dyn. Syst.}, 11(1):261--284, 2012.

\bibitem{burke2009swift}
John Burke, S~M Houghton, and Edgar Knobloch.
\newblock Swift-{H}ohenberg equation with broken reflection symmetry.
\newblock {\em Phys. Rev. E}, 80(3):036202, 2009.

\bibitem{houghton2011swift}
SM~Houghton and Edgar Knobloch.
\newblock Swift-{H}ohenberg equation with broken cubic-quintic nonlinearity.
\newblock {\em Phys. Rev. E}, 84(1):016204, 2011.

\bibitem{mercader2013travelling}
Isabel Mercader, Oriol Batiste, Arantxa Alonso, and Edgar Knobloch.
\newblock Travelling convectons in binary fluid convection.
\newblock {\em J. Fluid {M}ech.}, 722:240--266, 2013.

\bibitem{doedel08auto-07p}
Eusebius Doedel, Alan~R Champneys, Thomas~F Fairgrieve, Yuri~A Kuznetsov,
  Björn Oldeman, R~Paffenroth, B~Sandstede, Xianjun Wang, and C~Zhang.
\newblock {\em AUTO-07P: Continuation and Bifurcation Software for Ordinary
  Differential Equations}, 2012.

\bibitem{uecker_numerical_2021}
Hannes Uecker.
\newblock {\em Numerical {Continuation} and {Bifurcation} in {Nonlinear}
  {PDEs}}.
\newblock SIAM, Philadelphia, PA, 2021.

\bibitem{makrides2014predicting}
Elizabeth Makrides and Bj{\"o}rn Sandstede.
\newblock Predicting the bifurcation structure of localized snaking patterns.
\newblock {\em Physica D}, 268:59--78, 2014.

\bibitem{uecker2009short}
Hannes Uecker.
\newblock A short ad hoc introduction to spectral methods for parabolic
  {P}{D}{E} and the {N}avier-{S}tokes equations.
\newblock {\em Summer School Modern Computational Science}, pages 169--209,
  2009.

\bibitem{nishiura2003dynamic}
Yasumasa Nishiura, Takashi Teramoto, and Kei-Ichi Ueda.
\newblock Dynamic transitions through scattors in dissipative systems.
\newblock {\em Chaos}, 13(3):962--972, 2003.

\bibitem{nishiura2005scattering}
Yasumasa Nishiura, Takashi Teramoto, and Kei-Ichi Ueda.
\newblock Scattering of traveling spots in dissipative systems.
\newblock {\em Chaos}, 15(4):047509, 2005.

\bibitem{burke2009multipulse}
John Burke and Edgar Knobloch.
\newblock Multipulse states in the {S}wift-{H}ohenberg equation.
\newblock In {\em Conference Publications}, volume 2009, page 109. American
  Institute of Mathematical Sciences, 2009.

\bibitem{beck2009snakes}
Margaret Beck, J{\"u}rgen Knobloch, David~JB Lloyd, Bj{\"o}rn Sandstede, and
  Thomas Wagenknecht.
\newblock Snakes, ladders, and isolas of localized patterns.
\newblock {\em SIAM J. Math. Anal.}, 41(3):936--972, 2009.

\bibitem{bykov1980bifurcations}
VV~Bykov.
\newblock Bifurcations of dynamical systems close to systems with a separatrix
  contour containing a saddle-focus.
\newblock {\em Methods of the Qualitative Theory of Differential Equations},
  pages 44--72, 1980.

\bibitem{bykov1993bifurcations}
VV~Bykov.
\newblock The bifurcations of separatrix contours and chaos.
\newblock {\em Physica D}, 62(1-4):290--299, 1993.

\bibitem{glendinning1986t}
Paul Glendinning and Colin Sparrow.
\newblock T-points: a codimension two heteroclinic bifurcation.
\newblock {\em J. Stat. Phys.}, 43:479--488, 1986.

\bibitem{ROMEO2003242}
M\'{o}nica~M. Romeo and Christopher K. R.~T. Jones.
\newblock The stability of traveling calcium pulses in a pancreatic acinar
  cell.
\newblock {\em Physica D}, 177(1):242--258, 2003.

\bibitem{moreno2022bifurcation}
Pablo Moreno-Spiegelberg, Andreu Arinyo-i Prats, Daniel Ruiz-Reyn{\'e}s,
  Manuel~A Matias, and Dami{\`a} Gomila.
\newblock Bifurcation structure of traveling pulses in type-{I} excitable
  media.
\newblock {\em Phys. Rev. E}, 106(3):034206, 2022.

\bibitem{coullet1987nature}
P~Coullet, C~Elphick, and D~Repaux.
\newblock Nature of spatial chaos.
\newblock {\em Phys. Rev. Lett.}, 58(5):431, 1987.

\bibitem{elphick1991interacting}
Christian Elphick, Glenn~R Ierley, Oded Regev, and Edward~A Spiegel.
\newblock Interacting localized structures with galilean invariance.
\newblock {\em Phys. Rev. A}, 44(2):1110--1122, 1991.

\bibitem{ei1994equation}
Shin-Ichiro Ei and Takao Ohta.
\newblock Equation of motion for interacting pulses.
\newblock {\em Phys. Rev. E}, 50(6):4672--4678, 1994.

\bibitem{balmforth1994chaotic}
Neil~J Balmforth, Glenn~R Ierley, and Edward~A Spiegel.
\newblock Chaotic pulse trains.
\newblock {\em SIAM J. Appl. Math.}, 54(5):1291--1334, 1994.

\bibitem{nishiura2022traveling}
Yasumasa Nishiura and Takeshi Watanabe.
\newblock Traveling pulses with oscillatory tails, figure-eight-like stack of
  isolas, and dynamics in heterogeneous media.
\newblock {\em Physica D}, 440:133448, 2022.

\end{thebibliography}

\end{document}